\documentclass[11pt]{amsart}
\usepackage{graphicx}
\usepackage{amssymb,amsmath,amsthm}
\usepackage{epstopdf}
\usepackage{color}
\usepackage{mdframed}
\usepackage{empheq}
\usepackage{bm}
\usepackage{hyperref}
\usepackage{mathabx} 
\usepackage{enumitem}
\usepackage{xcolor}
\usepackage[linesnumbered,ruled,vlined]{algorithm2e}
\usepackage{subcaption}

\newtheorem{theorem}{Theorem}

\newtheorem{remark}[theorem]{Remark}

\makeatletter
\renewcommand\l@paragraph[2]{}
\renewcommand\l@subparagraph[2]{}
\makeatother
\title[Glasso Load and Wind Power]{GLASSO Model for Electric Load and Wind Power and Monte Carlo Scenario Generation}
\author{Ren\'e Carmona \& Xinshuo Yang}
\address{Department of Operations Research \& Financial Engineering\\
Princeton University\\
Princeton, NJ 08544, USA}
\date{February 20, 2021}                   

\begin{document}
\maketitle
\hfill

\begin{abstract}

For the purpose of Monte Carlo scenario generation, we propose a graphical model for the joint distribution of wind power and electricity demand in a given region. To conform with the practice in the electric power industry, we assume that point forecasts are provided exogenously, and concentrate on the modeling of the deviations from these forecasts instead of modeling the actual quantities of interest. We find that the marginal distributions of these deviations can have heavy tails, feature which we need to handle before fitting a graphical Gaussian model to the data. We estimate covariance and precision matrices using an extension of the graphical LASSO  procedure which allows us to identify temporal and geographical (conditional) dependencies in the form of separate dependence graphs.
We implement our algorithm on data publicly available for the Texas grid as managed by ERCOT, and we confirm that the geographical dependencies identified by the algorithm are consistent with the geographical relative locations of the zones over which the data were collected.  
\end{abstract}

\emph{Keywords:} Load, Wind power, Monte Carlo simulations, Graphical LASSO.

\section{\textbf{Introduction}}
\label{se:introduction}
Nowadays, the management of modern power grids relies heavily on large scale optimization programs responsible for unit commitment and economic dispatch. Some of the inputs of these programs require forecasts, and planning is often organized around attempt to feed these programs with scenarios, whether these scenarios are expected to be generic, desirable, or rare and to be protected against. We base our generation of such scenarios on scientifically rigorous principles and on sound statistical analyses. Also, we rely on work which has already been done to evaluate the \emph{quality} of the scenarios so generated. See for example \cite{PinsonGirard} for the analysis of wind scenarios.

Given the increased penetration of renewable energy production sources like solar and wind, the behavior of unit commitment and economic dispatch engines  has become a source of great concern. Analysts in the power generation industry quickly understood that they needed to account for highly uncertain and volatile inputs. As a result, they accepted the fact that they have to deal with probabilistic forecasts of wind, solar and even loads in order to understand the risks associated with the ever changing conditions under which they need to plan for systems operations. Engineers and applied mathematicians developed a wealth of tools to produce forecasts for loads and wind and solar power, and the industry adopted them as fundamental components of their operations. Historically, the wast majority of forecasts used by system operators were point forecasts. Unfortunately, because of their deterministic nature, these forecasts cannot easily account on their own for the variability of the actual future realizations of these variables, especially when they are as unpredictable and volatile as solar and wind power. Then came probabilistic forecasts. Even though they provide a form of quantification of the variability of the variables, and hence the reliability of the point forecasts, they are limited in their ability to capture temporal and spatial dependencies.

Henceforth, a large number of studies were devoted to the development of algorithms to generate statistical scenarios from probabilistic forecasts. For wind power, we refer to the fundamental paper \cite{Pinson_et_al_1} and the references therein, and to \cite{WatsonWets} for a recent \emph{state of the art} on the subject with a clear goal of providing inputs for unit commitment and economic dispatch platforms.

Very rarely were spatial and temporal correlations included in the same model, and when it was done, it was typically for one variable only. See for example \cite{Pinson_et_al_2} for an example dealing with wind. One of the main thrust of our paper is to propose algorithms to remedy this shortcoming by allowing temporal and spacial dependencies to be captured for wind power and loads simultaneously. We shall include solar in our algorithms in a forthcoming study. Modeling the random behavior of the future values of a variable, say wind, is commonly done on the basis of existing forecasts obtained exogenously. See for example \cite{wind_power}. This trend has become a standard in the industry, and we shall follow this time honored convention in the present paper.

This paper is a contribution to the literature on the modeling of the stochastic behavior of these quantities, and the development of simulation algorithms capable to deliver Monte Carlo scenarios for their future time evolution. 
While often touted in the published literature,  the analysis of fully correlated models is very rarely pushed beyond the level of ideas and discussions, and even more rarely implemented by means of simulation algorithms. See nevertheless the inclusion of time correlation for one single variable, e.g. wind, in \cite{Morales_et_al}.  The novelty of our approach has several prongs. First, we allow for the marginal distributions of the hourly zonal load and wind power variables to have heavy tails. So the first step is to check for the presence of heavy tails, and when found, fit a Generalized Pareto Distribution (GPD) to the marginal and use it to process the data to give them Gaussian marginals. We explain and illustrate the pros and the cons of this step with simulation illustrations on days for which the forecasts overestimate and underestimate the actual future values. In some respect, our approach is reminiscent of the Gaussian copula approach as advocated for example in \cite{Pinson_et_al_1}.
Another major feature of our model and its implementations is to be able to be used when the number of historical observations used to fit the model is small compared to the dimension of the state variables we model and simulate. In order to overcome the possible singularity of the empirical covariance matrix estimates, we use an approach based on graphical Gaussian models estimated by a form of the graphical lasso algorithm {\tt glasso}. In this way, we obtain reasonable proxies for the precision matrices and work with Gaussian models which can be simulated. As a further step to reduce the number of degrees of freedom we use the {\tt gemini} search for precision matrices in the form of tensor products to disentangle the contributions of the spatial and temporal components to the correlations introduced in \cite{Zhou}.

The paper is organized as follows: in Section \ref{se:set-up} we describe the data used for the analysis. The following Section \ref{se:loads} is devoted to the analysis of the demand for electric power in the four zones identified by the Texas ISO providing the data. The fitted model is the basis for the Monte Carlo simulation engine whose performance is demonstrated by examples of batches of scenarios. This is the main section of the paper. A similar analysis and performance demonstration is provided in Section \ref{se:wind} for the wind power generation in the five zone for which the ERCOT website provides data. Both load and wind generation are modeled together in Section \ref{se:altogether}. The main take-away from this section is that, despite the effort to model them both jointly, we could not find significant correlations them.
We conclude with a short recall of the main results of the paper in Section \ref{se:conclusion}.

\section{\textbf{Problem Set-Up and Description of the Data}}
\label{se:set-up}

\subsection{Case Study Data}
\label{sub:data}
As rehashed by all the news outlets reporting on the recent topsy-turvy weather conditions caused by the polar vortex, and the damages caused by the unusually cold temperatures and snow storms in parts of the US, Texas is an \emph{energy island}. For this reason, it is a very convenient testbed for quantitative analyses of a regional electric grid. The Electric Reliability Council of Texas (ERCOT) manages approximately $90\%$ of the Texas' electric load. It is the Independent System Operator (ISO) for the region. It  schedules power on an electric grid with very few interstate connections. It also performs financial settlements for the competitive wholesale of power electricity. ERCOT's web site provides a wealth of data.
Here is what we shall use in the present case study.

\vskip 6pt
The historical data  we secured covers the period 2018-01-01 to 2019-12-31. 
\vskip 4pt
For each day $d$, for each hour of the day $h$, and for each load zone $z\in\{\text{West}, \text{North}, \text{South}, \text{Houston}\}$, the load data set gives a numerical value for the aggregate load over the zone $z$ on day $d$ during the hour $h$. The right most pane of Figure \ref{fi:load_glasso} shows these four regions on a Texas map.
For each day $d$, for each hour of the day $h$, and for each load zone $z$, the load forecast data set gives a  $N_{lag}$ numerical vector of the $N_{lag}=24$ hours ahead of $(d,h)$ for which we have a point forecast of the load in that zone. 

\vskip 4pt
Similarly, for each day and hour of the day $(d,h)$ of the same period,  and for each wind zone $z\in\{\text{West}, \text{North}, \text{South}, \text{Coastal}, \text{Panhandle}\}$, the wind power data set gives a numerical value for the aggregate power generated by the wind farms located in the zone $z$ during the hour $h$ on day $d$. Notice that the wind zones are not in the same number as the load zones. Moreover, even when they have the same name, they do not coincide. This is very unfortunate for the analysis. The right most pane of Figure \ref{fi:wind_glasso} shows these five regions on a Texas map.
For each day and hour of the day $(d,h)$ and for each wind zone $z$,  
the wind forecast data set gives a  $N_{lag}$ numerical vector with $N_{lag}=24$ hours ahead of $(d,h)$ for which we have a point forecast of the wind power in that zone.

\begin{remark}
We conducted the same theoretical and numerical analyses for New York state data as available from the NY-ISO web site and we found that the results are very similar. Even though the structure of the data is different, for example the forecasts are not generated on a rolling basis like for ERCOT, we found the same stability and consistency between the spatial correlation graphs produced by the model and the geographical locations of the zones from which the data were collected. 
\end{remark}

\section{\textbf{Modeling \& Simulating Electricity Loads}}
\label{se:loads}

We first tackle the challenge of the design of the Monte Carlo scenario simulation engine for the loads over the region.

\subsection{\textbf{Strategy}}
As explained in our discussion in Section \ref{se:set-up}, we model the deviations (which are often called errors in the literature on the subject) defined as
\begin{center} 
\emph{load deviations} = \emph{load actuals} – \emph{load forecasts}.
\end{center}
Recall that the historical data  we secured covers the period 2018-01-01 to 2019-12-31. For each day and hour of the day $(d,h)$ of this period, for each load zone $z$, we read the vector of the $N_{lag}=24$ load point forecasts from the load forecast data set, and we subtract each lag $\ell\in\{0,1,\cdots,23\}$ point forecast from the actual load found in the load data set for $\ell$ hours past $(d,h)$.
This gives what we call the load deviation, the data set of all deviations having the same structure as the data set of load forecasts. 

\vskip 12pt\noindent
1)  For each load zone $z\in\{\text{West}, \text{North}, \text{South}, \text{Houston}\}$ and for each lag $\ell\in\{0,1,\cdots,23\}$ the load deviations form an hourly time series of length $N=17520$, the number of hours in the two years covered by our data sets. The first step of our analysis is to remove a trend and a seasonal component by standard regression techniques, leaving us with a time series of deviation remainders which we assume to be \emph{stationary}. Our next step is to analyze the marginal distributions of these $N^L_{zone}\times N_{lag}=4\times 24=96$ univariate time series.

\begin{figure}[h]
\centerline{
\includegraphics[width=4cm,height=4cm]{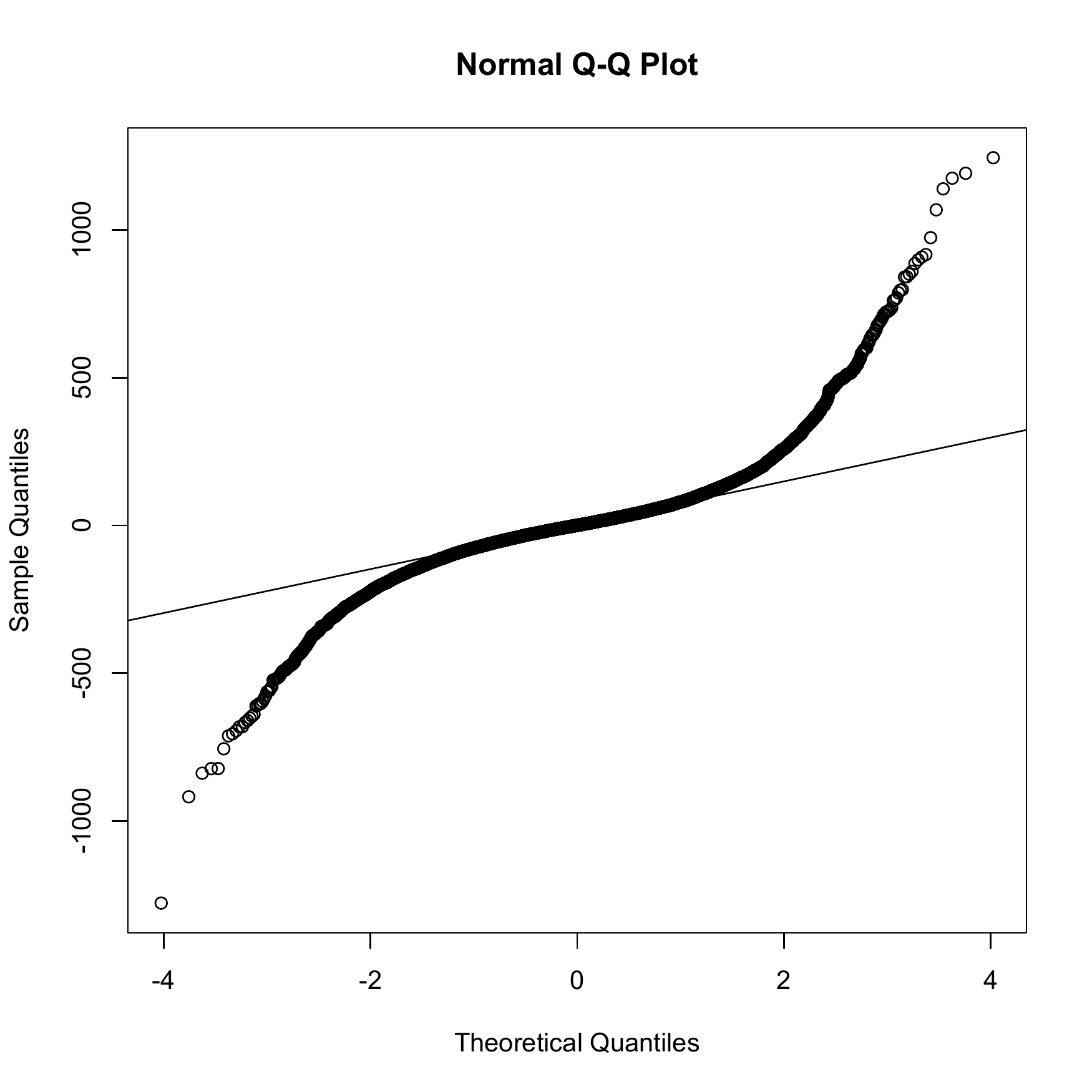}
\hskip 2pt
\includegraphics[width=4cm,height=4cm]{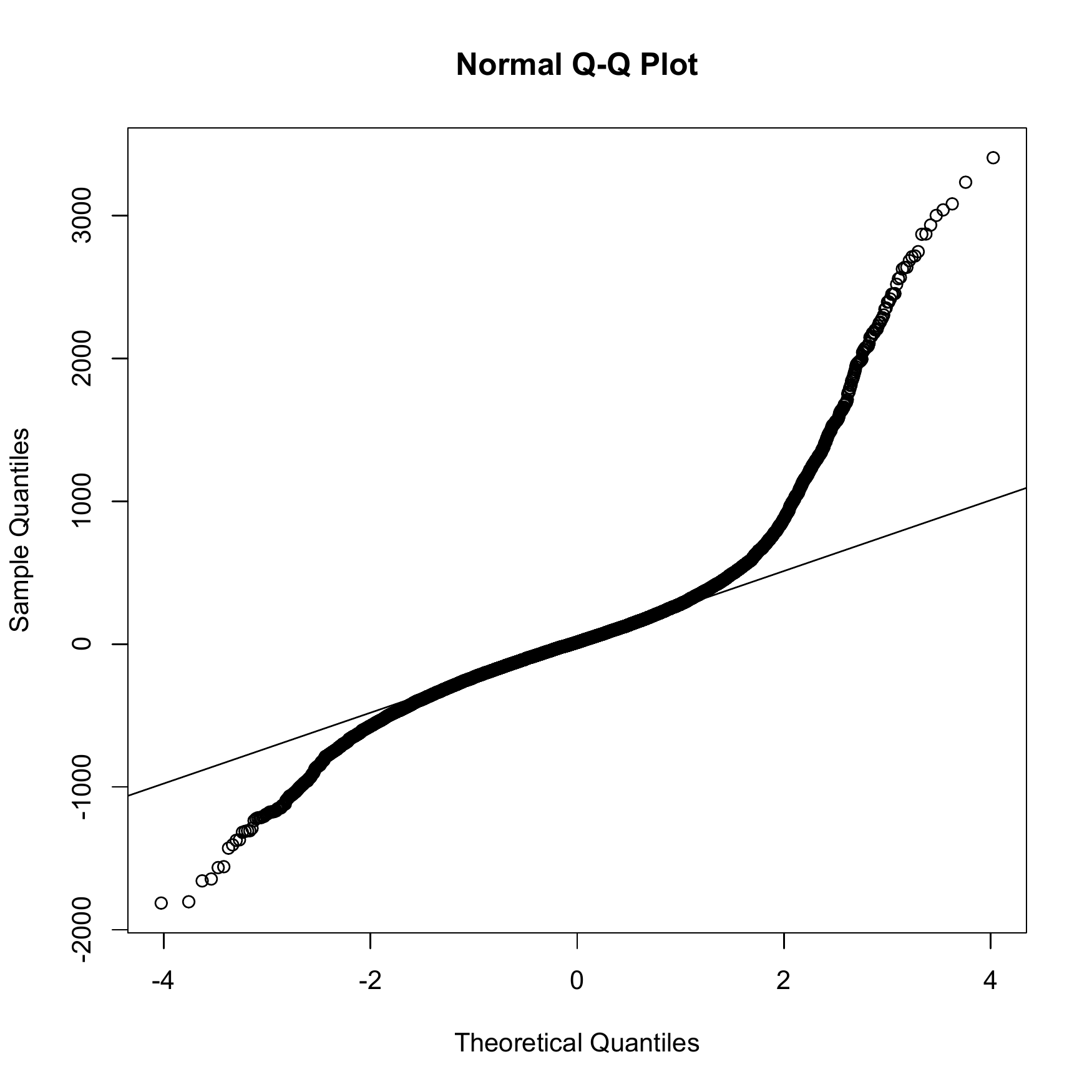}
\hskip 2pt
\includegraphics[width=4cm,height=4cm]{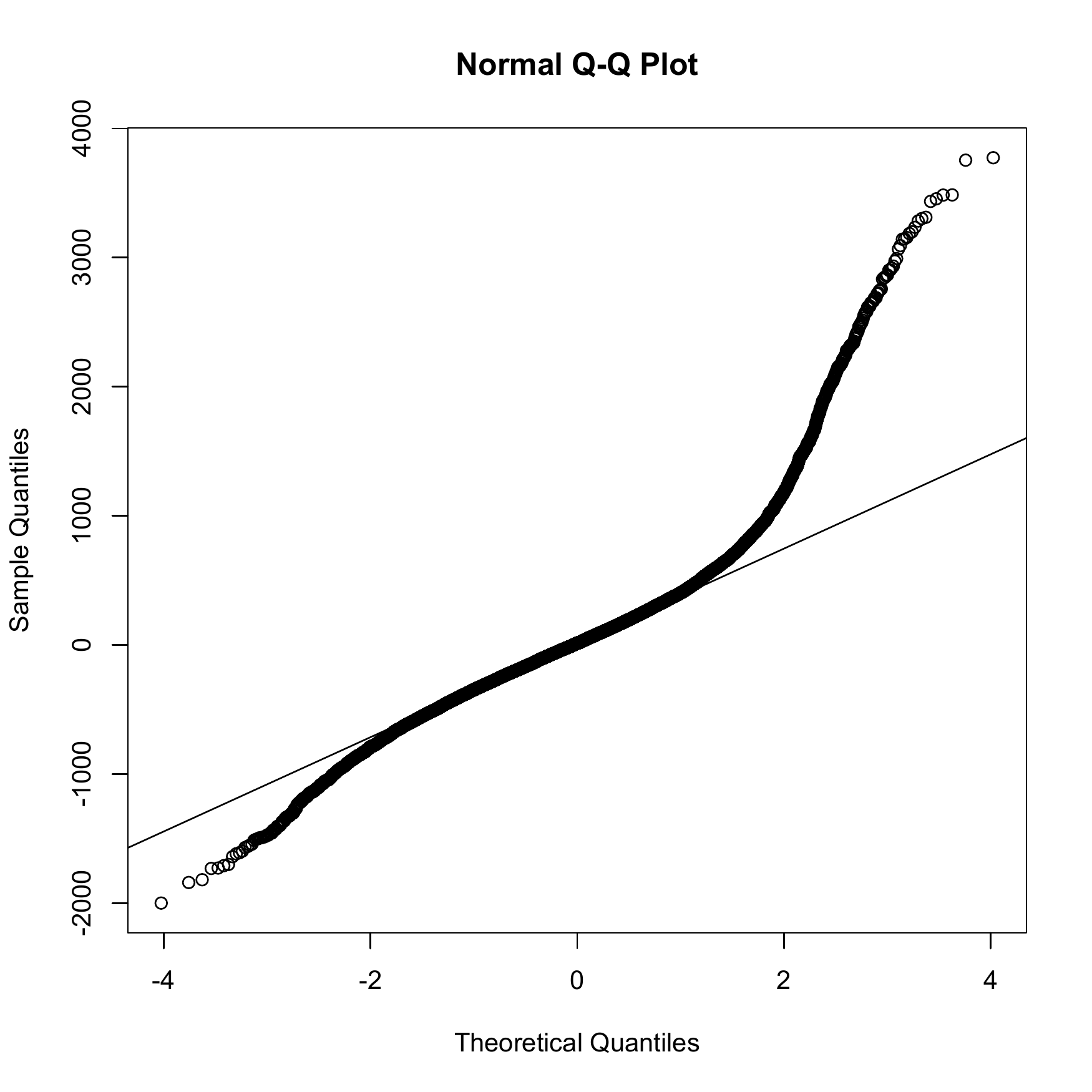}
\hskip 2pt
\includegraphics[width=4cm,height=4cm]{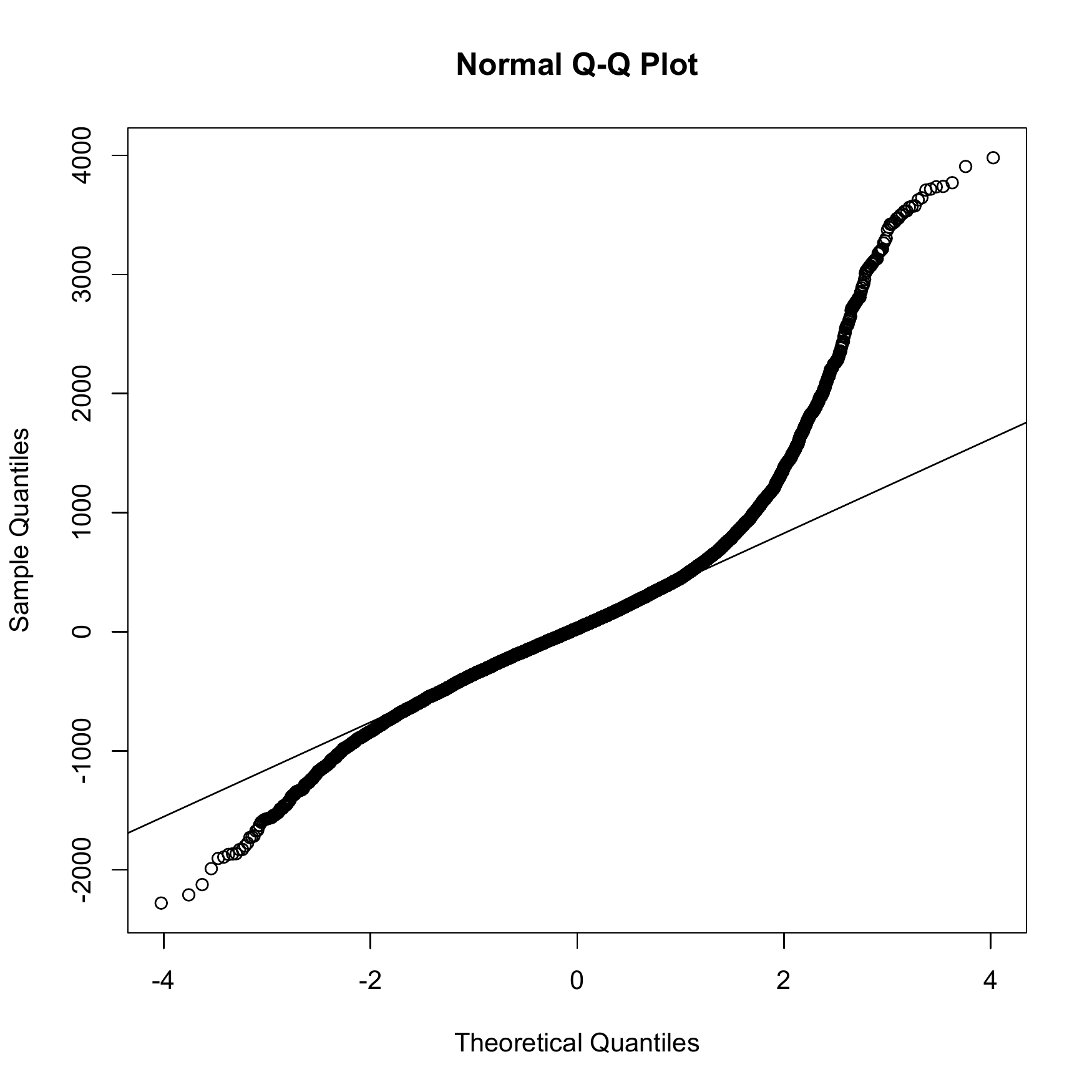}
}
\caption{Normal Q-Q plots for load deviation remainders in the $z=\text{WEST}$ region and lags $\ell=1, 4, 12$ and $\ell=23$ hours ahead.}
\label{fi:loads_qqnorms}
\end{figure}

\vskip 4pt\noindent
2) For each load zone $z$ and for each lag $\ell\in\{0,1,\cdots,23\}$ we produce a Q-Q plot of the empirical quantiles of the load time series against the theoretical quantiles of the standard Gaussian distribution $N(0,1)$. Figure \ref{fi:loads_qqnorms} give a sample of $4$ of these Q-Q plots. They are typical of what we found throughout. This is clear evidence that the marginal distributions of the load deviation remainders have \emph{heavy tails}, justifying our next step. Notice also that these plots show that in these four cases, as well in the vast majority of the cases, the upper tail is heavier than the lower tail of the distribution.

\vskip 4pt\noindent
3) For each load zone $z$ and for each lag $\ell\in\{0,1,\cdots,23\}$ we fit a Generalized Pareto Distribution (GPD), see for example \cite[Chapter 2]{Carmona_SAFD}, to the  load deviations. We denote such a distribution as $G^L_{z,\ell}$ for the sake of later reference. Throughout this work, in order to manipulate GPD distributions, we use functions of the {\tt R} library {\tt Rsafd} \cite{rsafd} developed for the analysis of the examples and problem sets of the book \cite{Carmona_SAFD}, which we wrapped into a Python package.

\begin{figure}[h]
\centerline{
\includegraphics[width=4cm,height=4cm]{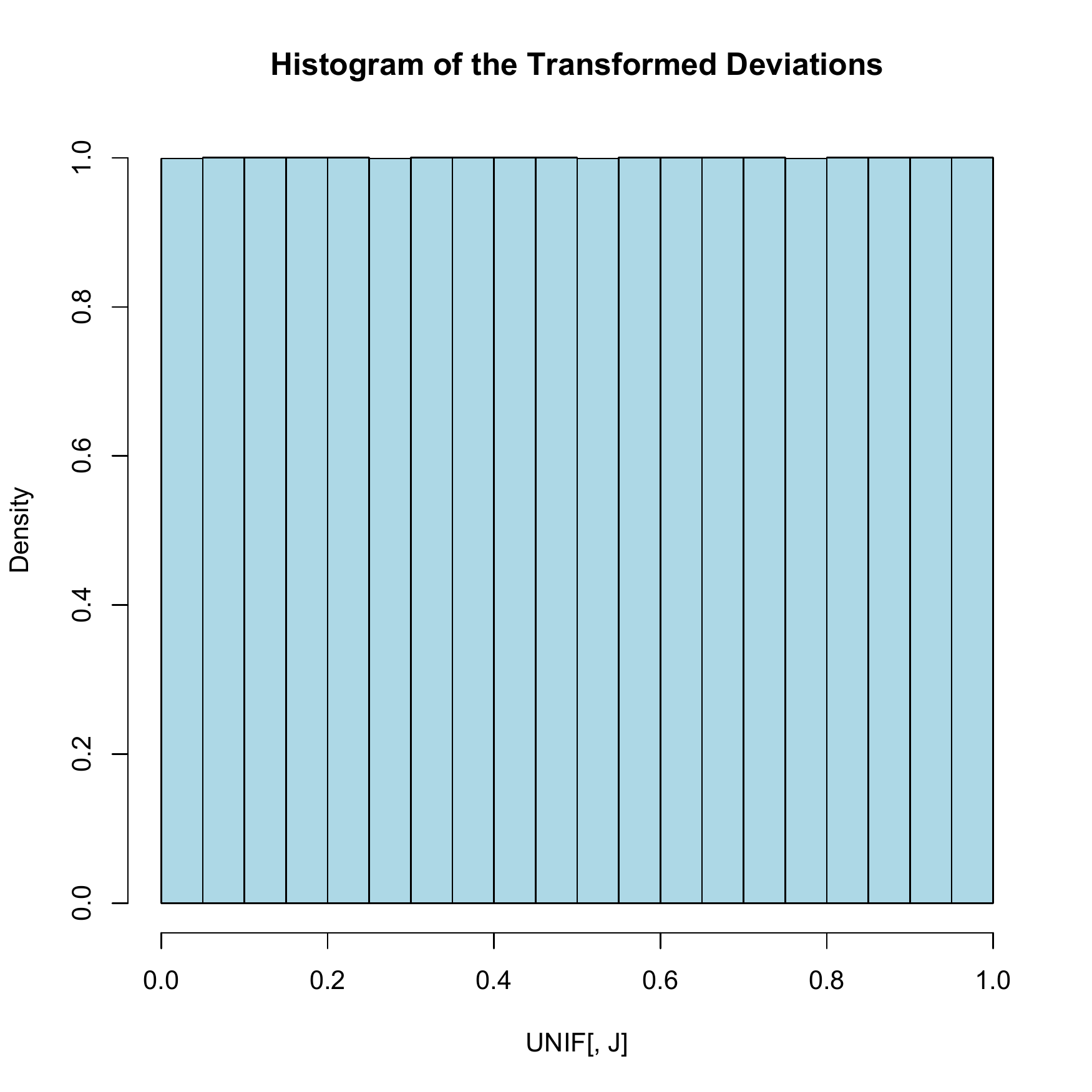}
\hskip 6pt
\includegraphics[width=4cm,height=4cm]{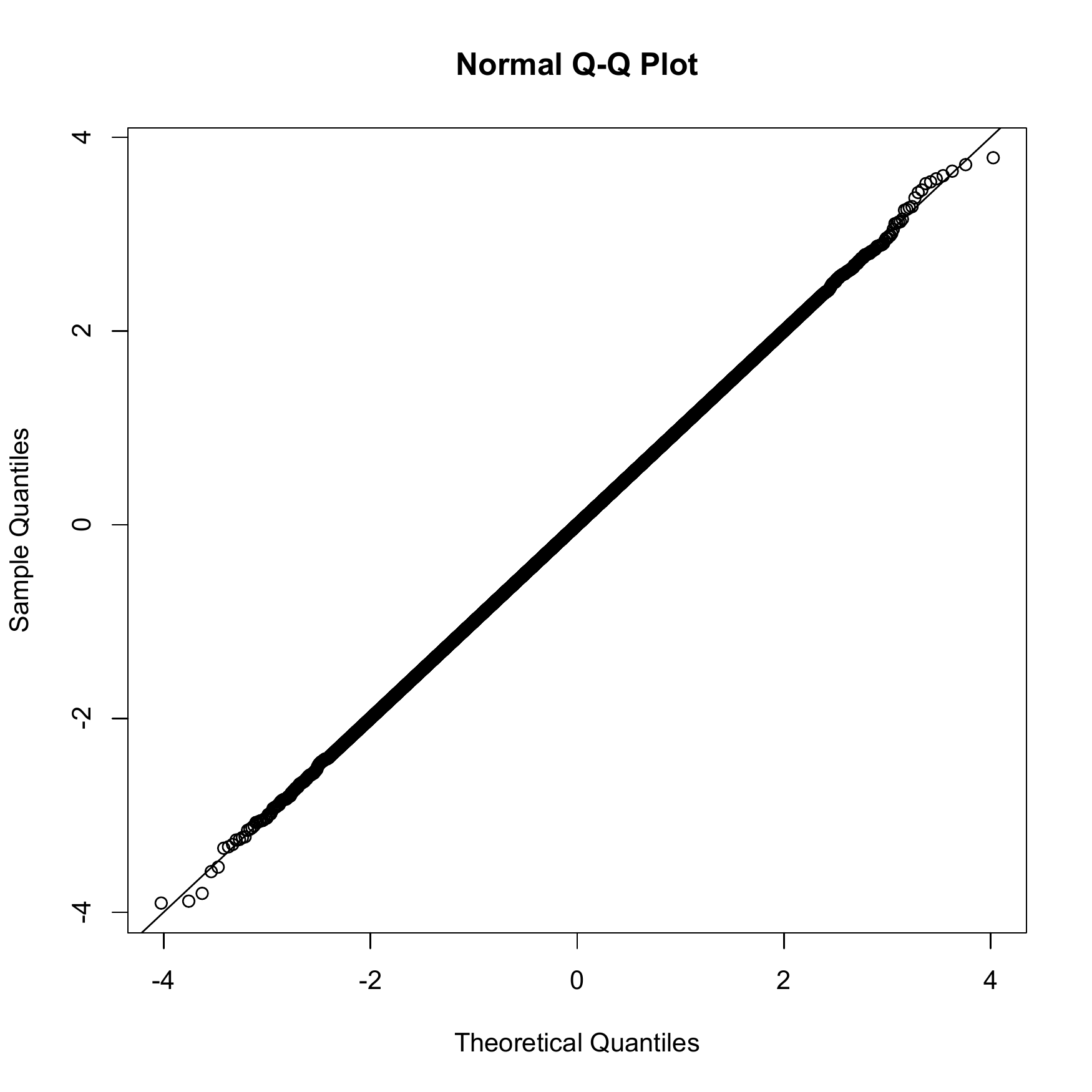}
}
\caption{Transformations of the load deviations in the WEST region $4$ hours ahead.}
\label{fi:load_tranform}
\end{figure}

\vskip 4pt\noindent
4) For each load zone $z$ and for each lag $\ell\in\{0,1,\cdots,23\}$ 
we transform the load deviation remainder time series into a uniform time series by computing the cumulative distribution function of the GPD just fitted on each of the entries of the time series. If the fitted GPD is a good fit, we should expect the marginal distribution of the new time series to be uniform. The histogram on the left pane of Figure \ref{fi:load_tranform}  is a testament to this claim.

\vskip 4pt\noindent
5) For each load zone $z$ and for each lag $\ell\in\{0,1,\cdots,23\}$ we transform the uniform time series just obtained into a time series with a standard Gaussian distribution $N(0,1)$. In order to do that, we just compute the quantile function of the standard Gaussian distribution on each of the entries of the uniform time series. The normal Q-Q plot on the right pane of Figure \ref{fi:load_tranform} gives a graphical evidence of the fact that the marginal distribution of the resulting time series is indeed a standard Gaussian distribution.

\vskip 4pt\noindent
6) At this stage of the analysis, for each day and hour of the day $(d,h)$, we have a $N^L_{zone}\times N_{lag}=4\times 24=96$ dimensional numerical vector, each component of this vector having a standard Gaussian distribution. While we do not know if this $96$-dimensional vector is actually \emph{jointly Gaussian}, we shall behave as if that was the case, and we proceed to fit a \emph{Gaussian Graphical Models} using {\tt LASSO} to estimate the precision and covariance matrices. The interested reader will find a readable introduction to Gaussian graphical models, including LASSO estimation in \cite[Chapter 11]{Wainwright}. Here we could use the {\tt R} package {\tt glasso} to fit the graphical model to our modified load data. However, we believe that we would miss the special structure of our $96$ dimensional vector which has a $4$-dimensional spatial component and a $24$-dimensional temporal component. Relying on the intuition developed for classical statistical analysis of variance with multiple factors, we would like to think of the $96\times 96$ covariance and precision matrices as  tensor products (also known as Kronecker products) of a $4\times 4$ covariance and precision matrices for the spatial component, and a $24\times 24$ covariance and precision matrices for the temporal component. One way to do just that is to use the {\tt GEMINI} algorithm proposed in \cite{Zhou}.

\begin{figure}[h]
\centerline{
\includegraphics[width=5cm,height=5cm]{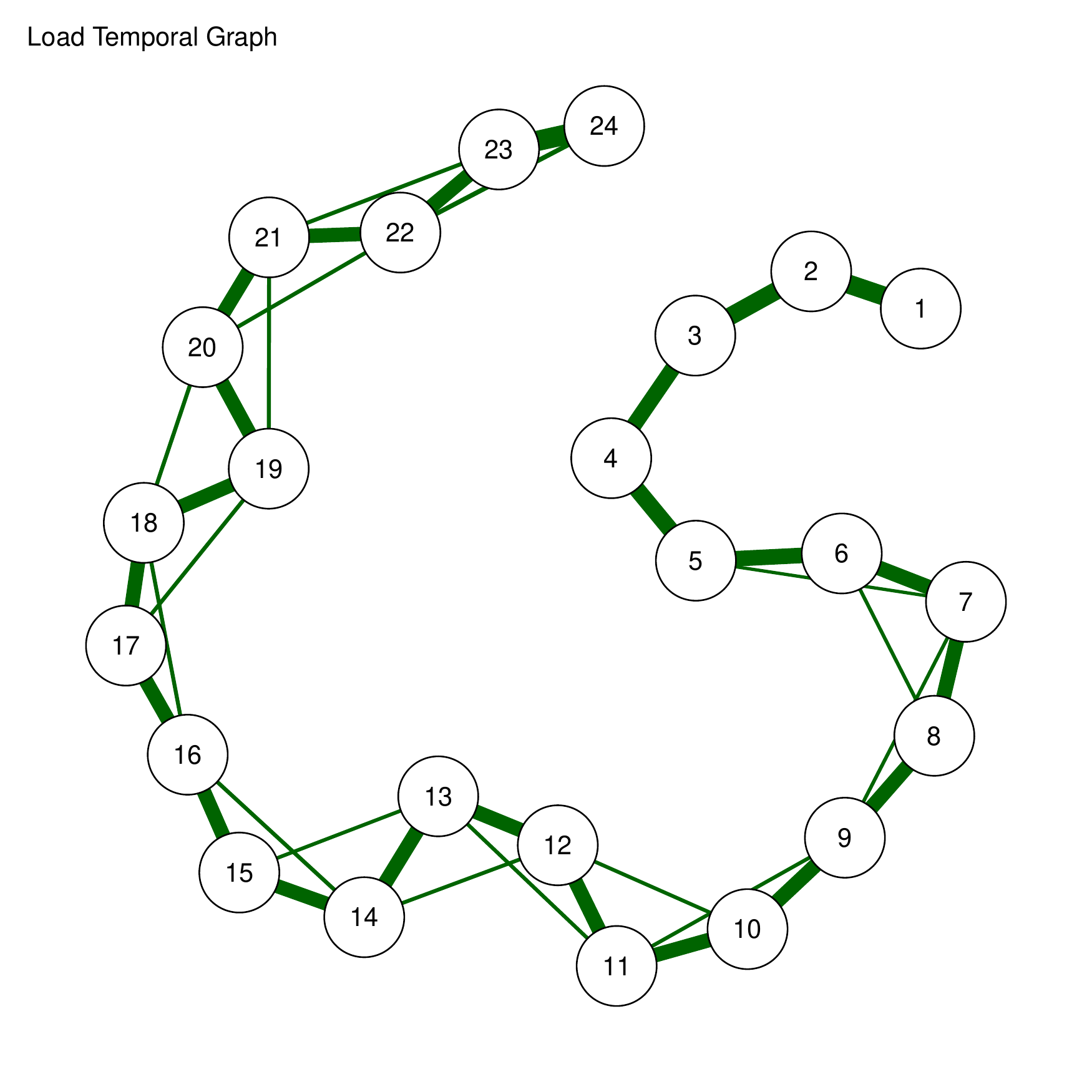}
\hskip 6pt
\includegraphics[width=5cm,height=5cm]{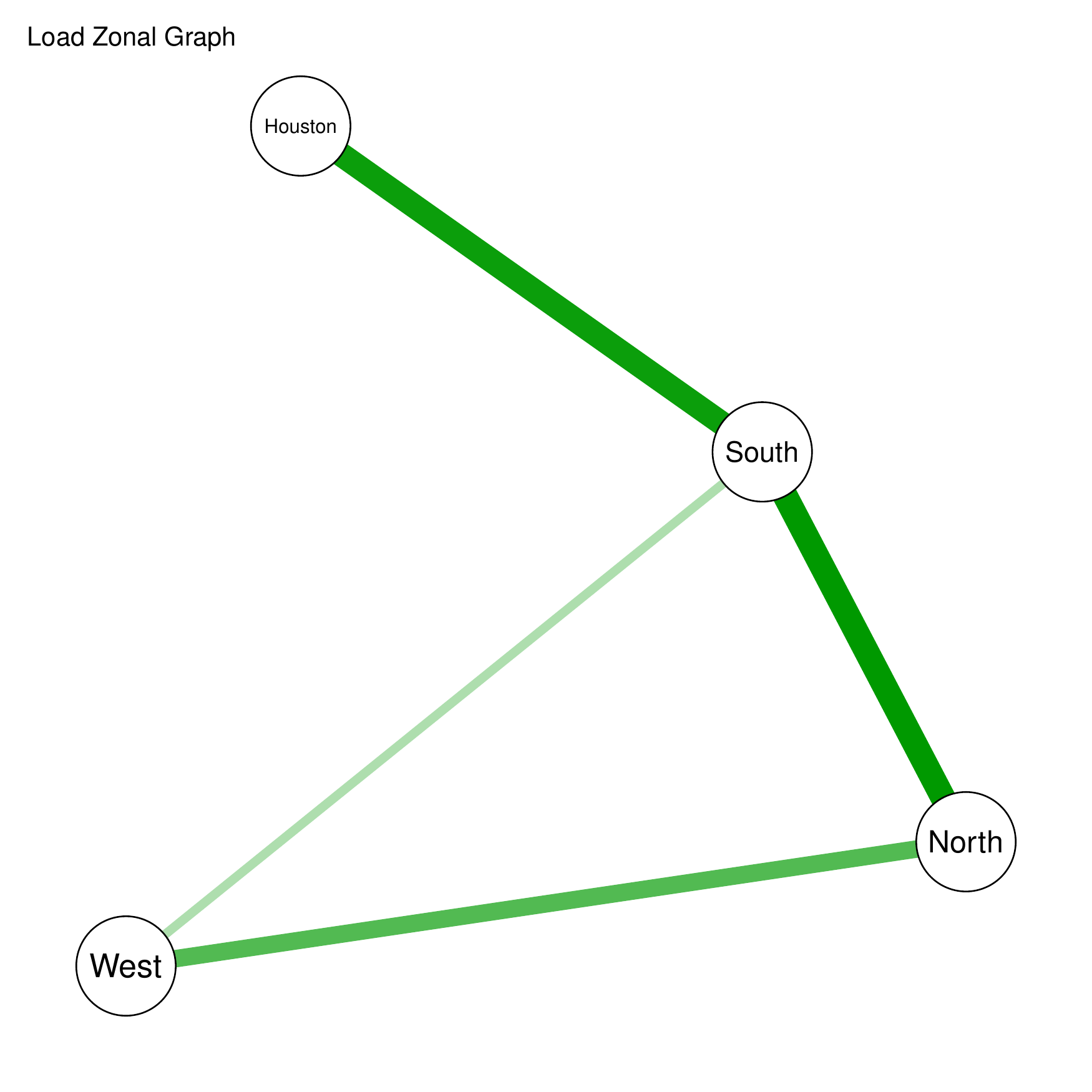}
\hskip 6pt
\includegraphics[width=7cm,height=5cm]{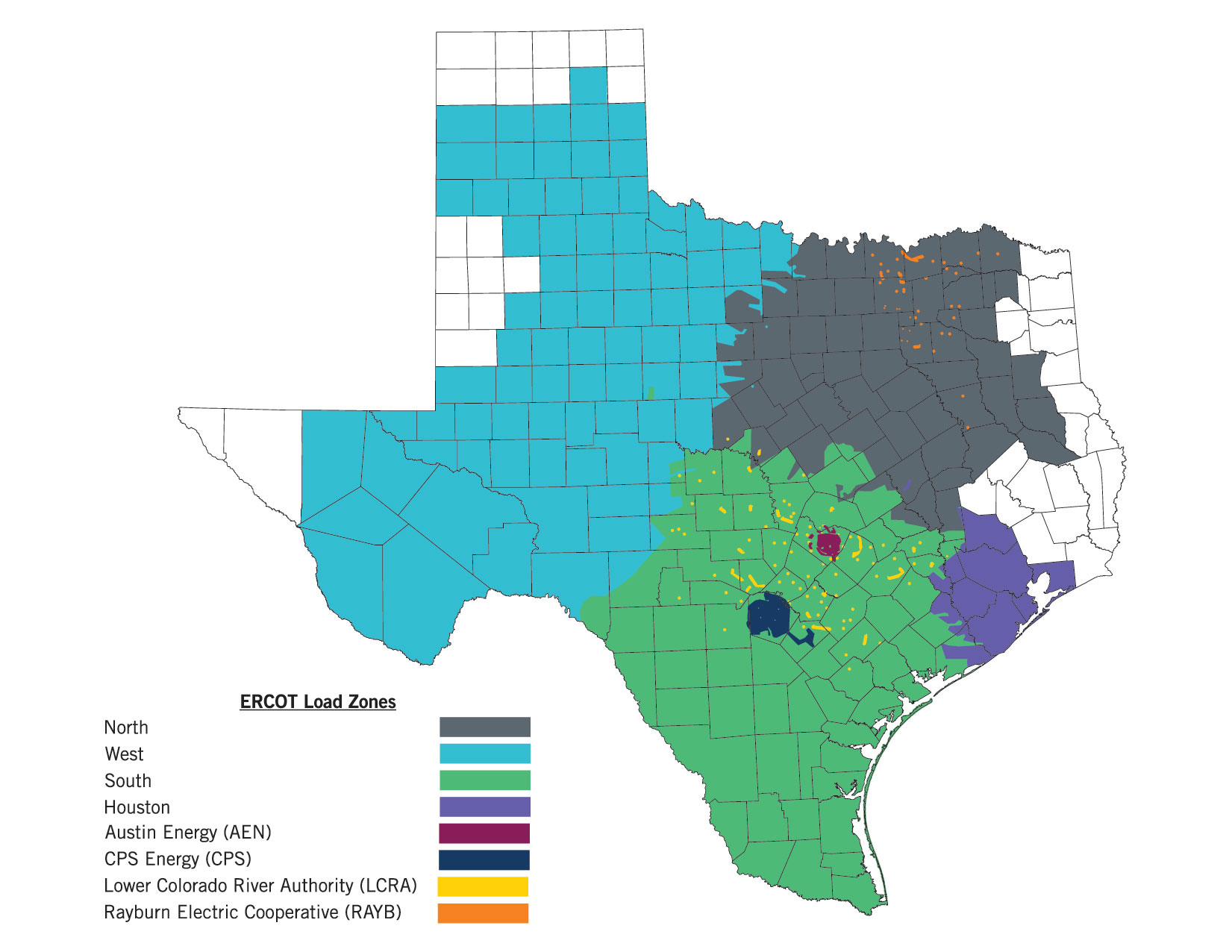}
}
\caption{Temporal component of the conditional correlation (left), spatial component of the conditional correlation (middle), geographical map of the four main load zones in ERCOT.}
\label{fi:load_glasso}
\end{figure}

\vskip 4pt
Figure \ref{fi:load_glasso} shows the graph of the conditional temporal dependence in the left pane, as well as the spatial conditional dependence component provided by the {\tt GEMINI} algorithm. 
\vskip 12pt

The graph of the temporal dependencies comprises two types of links. The thick links form a perfectly ordered linear chain. It points clearly at some form of Markovian structure. However, the thinner links show a residual dependence for hours two lags apart, which is quite intuitive. The edges, as well as their strengths in the graph of the spatial dependencies form a pattern consistent with the contiguity of the load regions as shown in the right pane of Figure \ref{fi:load_glasso}. Notice that the plotting package used to produce these plots labeled the lags from $1$ to $24$ instead of from $0$ to $23$ as we referred to then earlier. 

\vskip 4pt\noindent
7)  The next step of the analysis is to generate Monte Carlo samples from the model. Given the covariance matrix estimated above, we can easily generate as many Monte Carlo samples from the $96$-dimensional mean zero Gaussian distribution with this covariance matrix. Computing the cumulative distribution function of the standard Gaussian distribution on each of the entries provides sample of $96$-dimensional random vectors with uniform marginals. Each of the $96$ components corresponds to a specific zone $z$ and a specific lag $\ell$ so if we now compute the quantile function of the GPD distribution $G^L_{z,\ell}$ fitted to the original load deviations for zone $z$ and lag $\ell$, we obtain samples of a $96$ dimensional random vector with marginal distributions originally fitted to the load deviation time series. At this stage, it is important to emphasize that these $96$ components depend upon each other in the way dictated by the covariance structure found in the estimate of the graphical Gaussian model. The astute reader will have certainly noticed that we are actually using a $96$-dimensional copula to build the dependencies between the $G^L_{z,\ell}$ marginals of this $96$-dimensional vector.

At this stage, it is very easy to complete the simulation cycle by adding back the trend and  seasonal components identified for each $(d,h)$ and each $(z,\ell)$, in which case we now have Monte Carlo samples from the deviations, and so, adding back the forecasts, we end up with Monte Carlo scenarios of the actual loads.

\subsection{\textbf{Implementation}}
\begin{figure}[h]
\centerline{
\includegraphics[width=8cm,height=6.5cm]{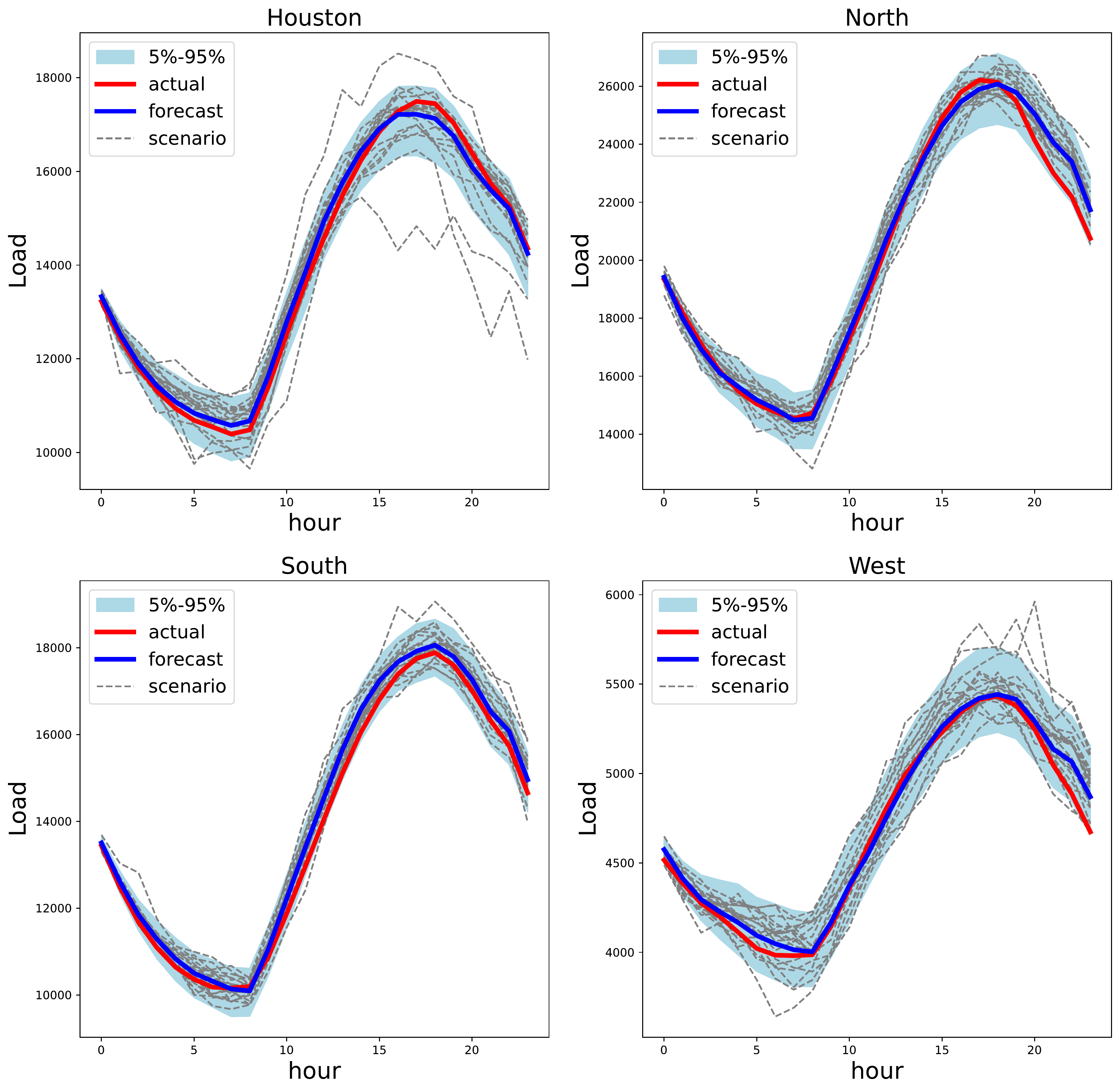}
\hskip 6pt
\includegraphics[width=8cm,height=6.5cm]{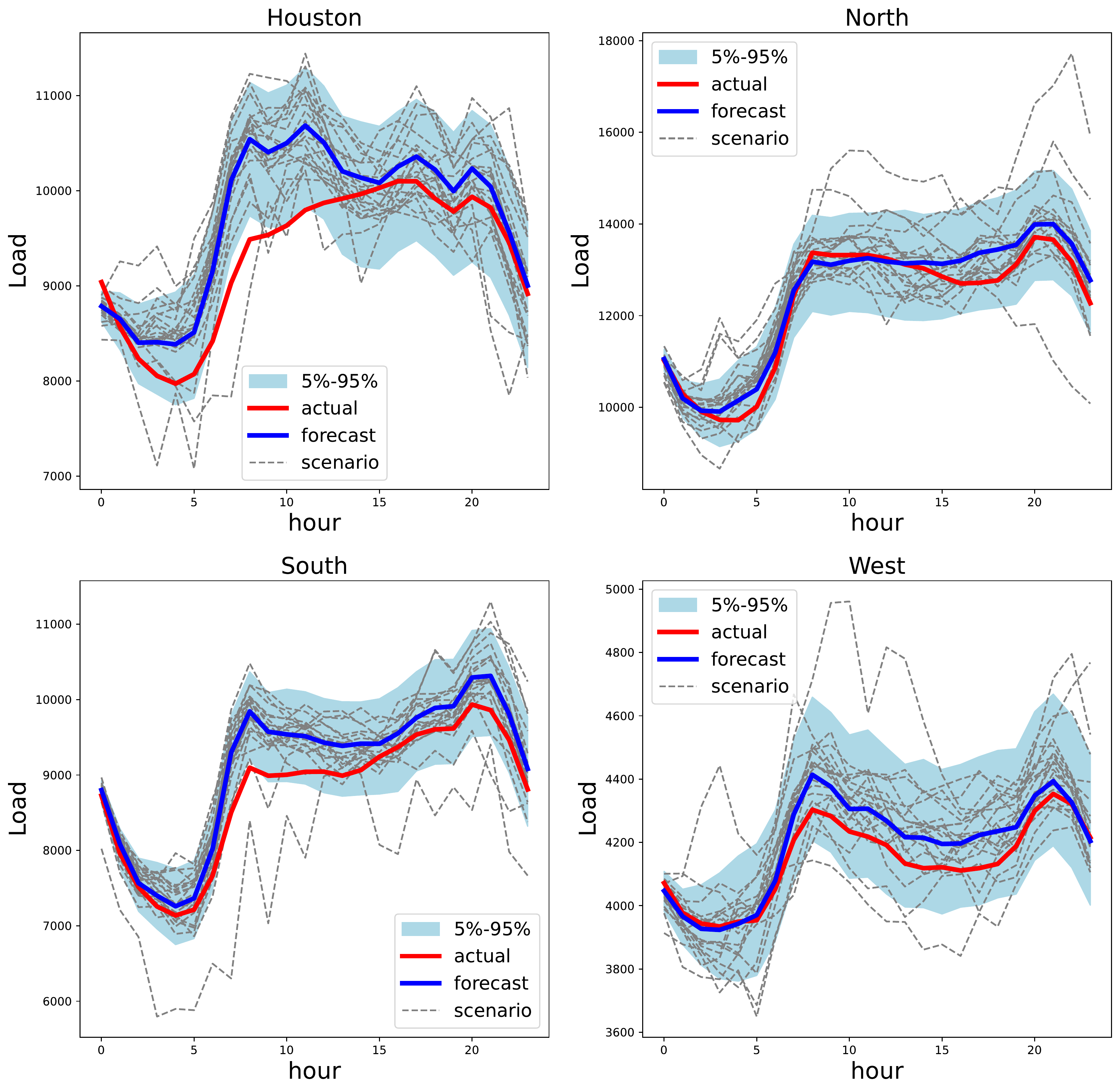}
}
\caption{Two examples of sets of $1000$ Monte Carlo scenarios for the next $24$ hours of the actual loads for July 1, 2018 (left)
and November 1, 2018 at midnight. The red line gives the actual loads observed over these $24$ hours, the blue line shows the point forecasts for these $24$ hours while the dashed lines are $20$ randomly selected scenarios. The gray bands are the trace left by the Monte Carlo scenarios after we remove the smaller $5\%$ and the larger $5\%$ of the bunch.}
\label{fi:load_scenarios}
\end{figure}

Figure \ref{fi:load_scenarios} gives an illustration of what the scenarios look like vis-a-vis the actual and forecasts loads in two specific choices of $(d,h)$. 

\begin{remark}
At this stage it is important to emphasize that because of the lack of enough historical data, we are working \emph{in sample} 
for the purpose of the above demonstration. In other words, our simulations use a form of \emph{crystal ball} since they use all the  observations in the data set to fit the model from which Monte Carlo simulations are generated. This shortcoming can easily be remedied if more historical data is available, and our experience is that the performance of the scenarios will not deteriorate significantly.
\end{remark}

\begin{remark}
Notice also that in order to have enough data at our disposal, we worked under the assumption that the load deviations have a single distribution, irrespective of the day of the year (whether $d$ is in the summer or the winter, or $\ldots$) or the time of the day determined by the hour $h$ and the lag $\ell$. Our reason for doing that is that we expect that the major trend and seasonal effects will be captured  by the individual trend and seasonal components estimated for the raw deviations, and that the remaining effects which would require working with conditional distributions instead of marginal distributions, are captured by the exogenously provided forecasts which are usually constructed from a large number of meteorological variables and Bayesian averaging of many forecasting models.
\end{remark}

\subsection{\textbf{The Impact of Fitting Heavy Tail Distributions}}
Here, we discuss the impact of our decision to check for the presence of heavy tails and to possibly fit generalized Pareto distributions to the marginal laws of the deviations. In order to do so, we provide a detailed examination of the smoothness of the Monte Carlo scenarios and the coverage provided by these scenarios.

\begin{figure}[h]
\centerline{
\includegraphics[width=12cm,height=6cm]{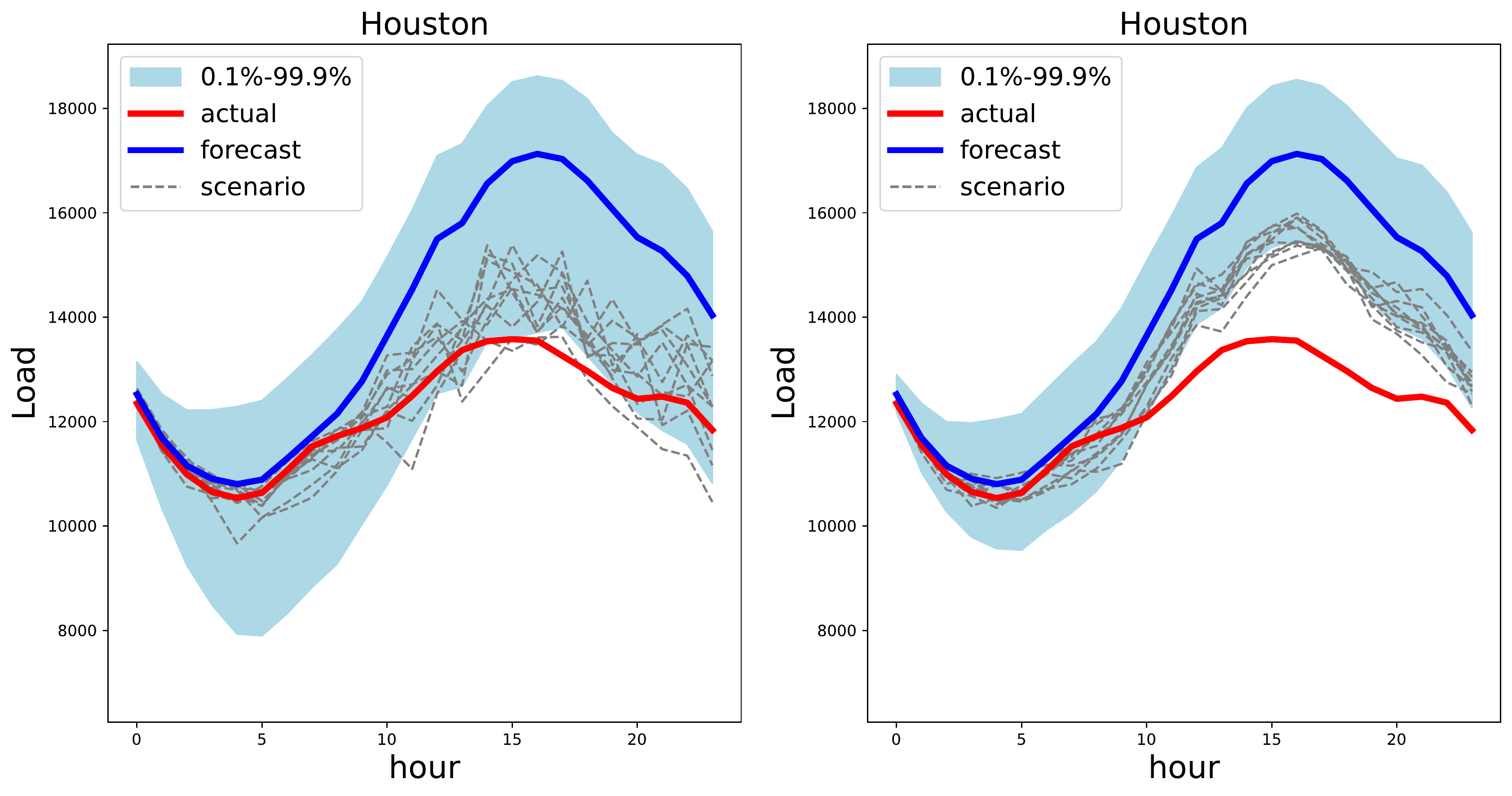}
}
\caption{Actual loads (red curves) and forecasts (blue curve) as well as a gray band covering $10,000$  Monte Carlo scenarios from the stochastic model, as well as a few sample scenarios (dashed lines) for the $24$ hours following June 25, 2018. The left pane was produced using the model described in the paper, including fitting (possibly) heavy tail distributions to the individual deviations, while the right pane was produced from a Gaussian model ignoring the possible presence of heavy tails in the distributions of the deviations.}
\label{fi:overestimation}
\end{figure}

\vskip 2pt
Repeated overestimation by the forecasts create instances of the negative values for the deviations, and if these instances are significant because of their sizes, the empirical distribution of the deviations can exhibit a heavy lower tail which cannot be accounted for by Gaussian distributions.  We illustrate  such an instance in Figure \ref{fi:overestimation} with the example of the Houston load zone on June 25, 2019. The left pane of the figure shows that the actual loads (red curve) hover around the lower boundary of the band created by the Monte Carlo scenarios. Recall that these scenarios are obtained by adding sample scenarios of the deviations to the forecasts (blue curve). However,  fitting a distribution with a heavy lower tail guarantees that the corresponding scenarios will cover the actuals values. The latter should be viewed as extremes, possibly rare events, but still in the realm of the model. The right pane shows the same plot produced from Monte Carlo scenarios from a Gaussian model ignoring the presence of heavy tails. While one could think that having a narrower coverage band is desirable, the fact that the actual values are so far from what the model considers to be reasonable is a major flaw of this form of the model.

\begin{figure}[h]
\centerline{
\includegraphics[width=12cm,height=6cm]{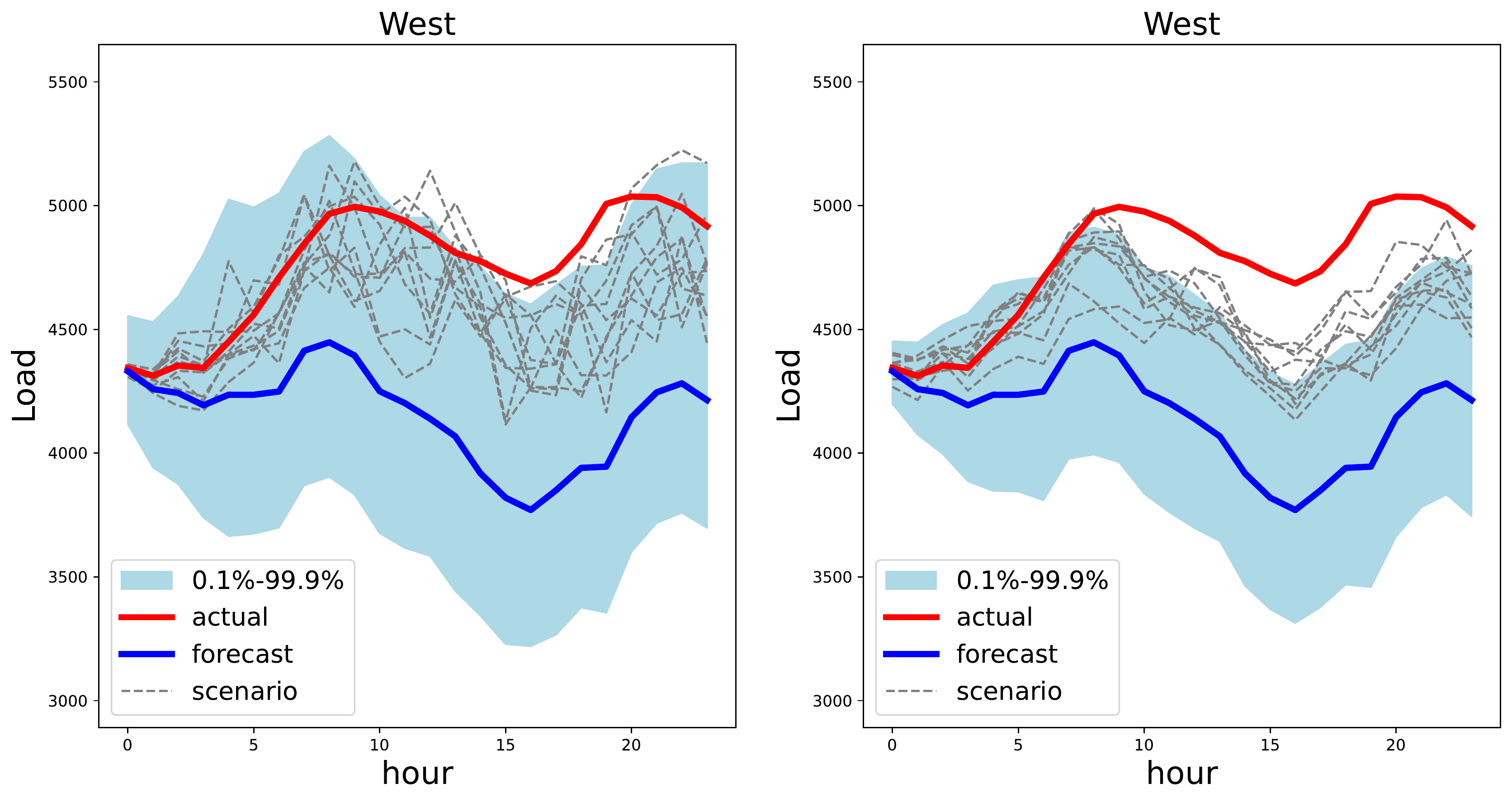}
}
\caption{Actual loads (red curves) and forecasts (blue curve) as well as a gray band covering $10,000$  Monte Carlo scenarios from the stochastic model, as well as a few sample scenarios (dashed lines) for the $24$ hours following December 28, 2018. The left pane was produced using the model described in the paper, including fitting (possibly) heavy tail distributions to the individual deviations, while the right pane was produced from a Gaussian model ignoring the possible presence of heavy tails in the distributions of the deviations.}
\label{fi:underestimation}
\end{figure}

In a similar fashion, we can discuss the case of underestimation by the forecasts. Clearly such instances can have dire consequences for the management of the electric grid, and they are feared by system operators. Figure \ref{fi:underestimation} provides an illustration which makes a resounding case for the use of heavy tail distributions in the model. We use the example of the load in the West region on December 28, 2018. With the same convention as before, the left pane of Figure \ref{fi:underestimation} shows that the actual loads for the next $24$ hours (red curve), while much higher than the forecasts (blue curve) are mostly covered by the scenarios. So even on the tail end of the possibilities given by the Monte Carlo simulations, it can still be expected to occur with significant probability and the operator of the grid should not dismiss such a pattern.
However, the right pane of Figure \ref{fi:underestimatiom} shows the result of the same Monte Carlo simulations when the model does not attempt to detect and fit heavy tail distributions. Clearly the curve of actual loads is far beyond the simulation band and an operator relying on such a model will consider the actual $24$ actual values as a possibility with zero likelihood.

The two instances discussed above clearly illustrates the pros of checking for the presence of heavy tails in the marginal distributions of the deviations, and when detected, of using generalized Pareto distributions to process the data. Still, a close look at the scenarios plotted in Figure \ref{fi:overestimatiom} and Figure \ref{fi:underestimatiom} shows that the scenarios in the left panes of these figures are not as smooth as those in the right panes. Beyond being less pleasing to the eye, this unavoidable side effect could possibly be regarded as a negative feature of the algorithm.

\section{\textbf{Modeling \& Simulating Wind Power}}
\label{se:wind}

This section will be shorter than the previous one. Indeed, we present the same analysis as for the loads, emphasizing only the significant differences when they are relevant.

\subsection{\textbf{Strategy}}
\begin{figure}[h]
\centerline{
\includegraphics[width=4cm,height=4cm]{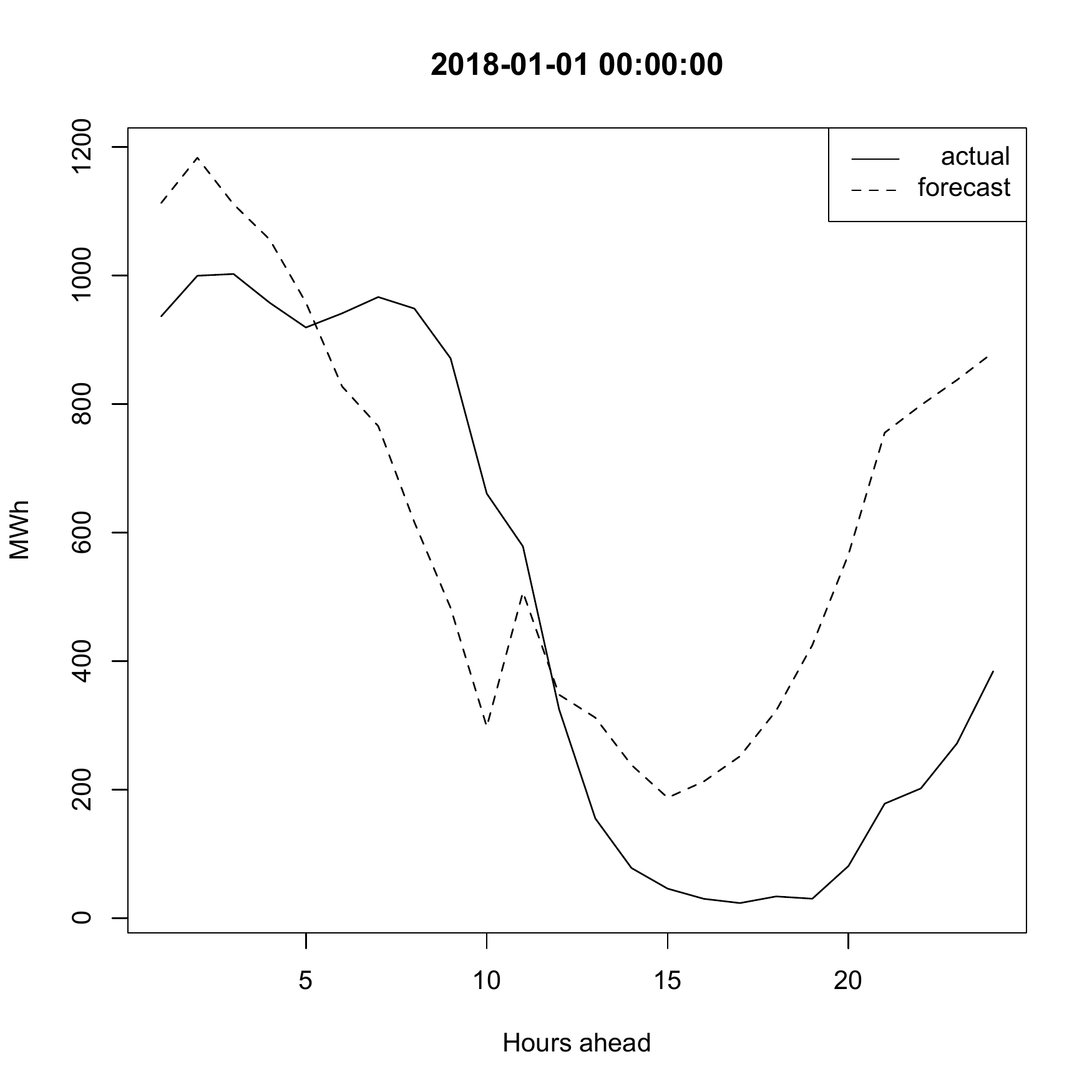}
\hskip 6pt
\includegraphics[width=4cm,height=4cm]{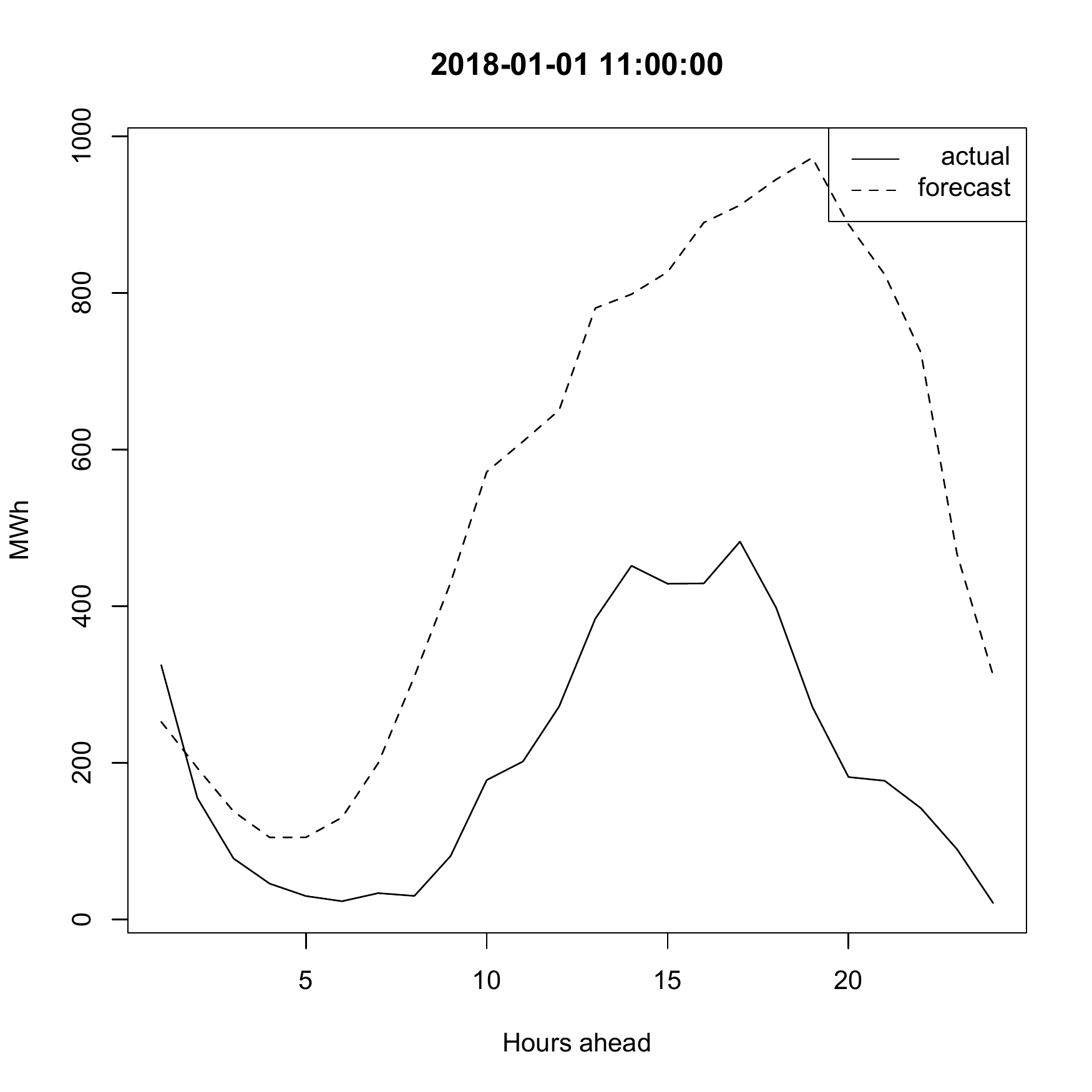}
\hskip 6pt
\includegraphics[width=4cm,height=4cm]{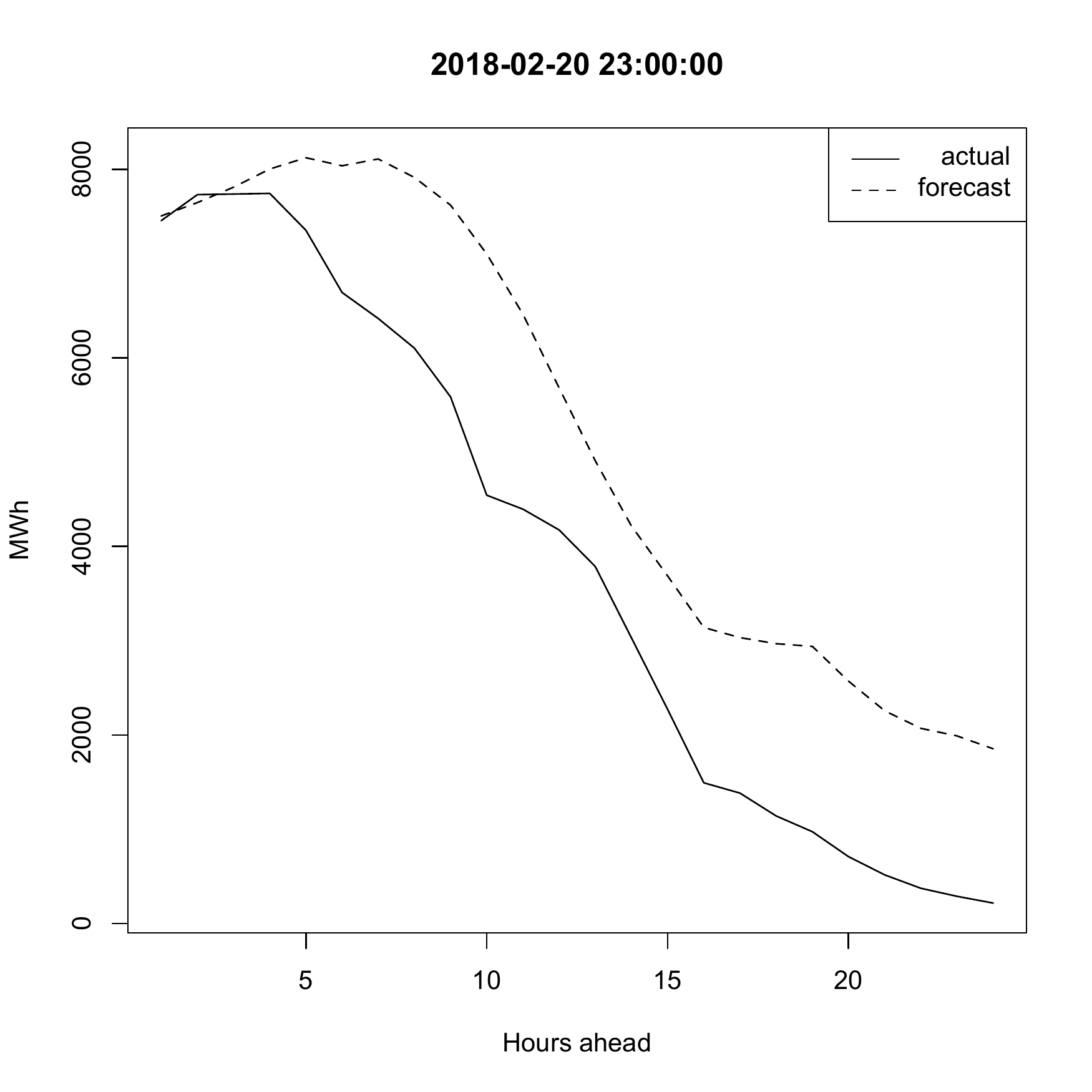}
\hskip 6pt
\includegraphics[width=4cm,height=4cm]{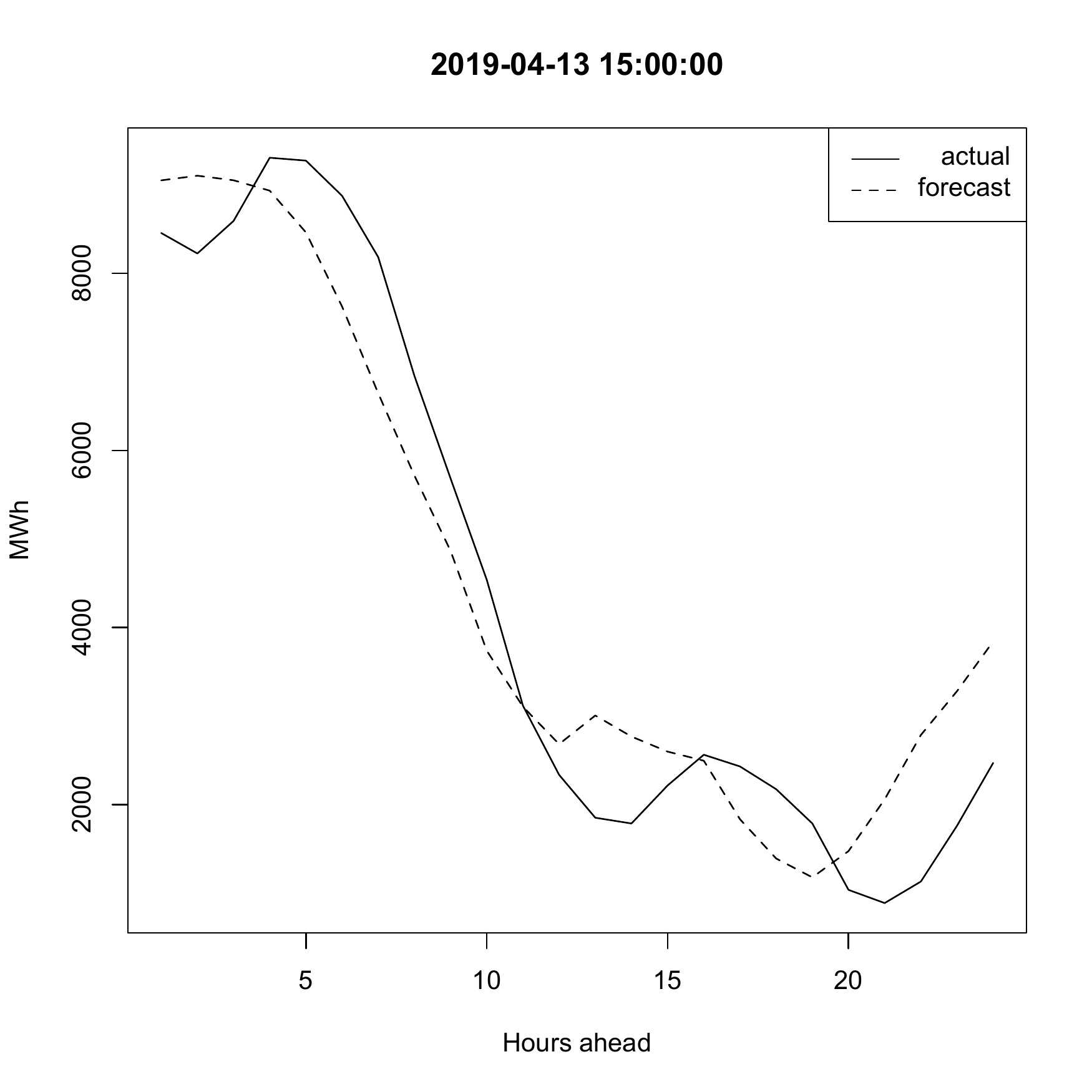}
}
\caption{Plots of the actual wind power and the forecasts against the lag  $\ell=1, \cdot, 24$ 
generated at  from left to right,  $(d,h)=2018-01-01 00:00:00$, $(d,h)=2018-01-01 11:00:00$, $(d,h)=2018-02-20 23:00:00$, and $(d,h)=2018-04-13 15:00:00$ in the $z=$WEST wind power zone.}
\label{fi:actual_forecast}
\end{figure}

As before, we introduce the \emph{deviations} which we define as
\begin{center} 
\emph{wind power deviations} = \emph{wind power actuals} – \emph{wind power forecasts}, 
\end{center}
and we prepare the data set of wind power deviations exactly as we did for the loads, for the same period ranging from 2018-01-01 and ending on 2019-12-31, the only noticeable difference being that we have now $N^W_{zone}=5$ zones which are now called West, North, South, Coastal, and Panhandle. We still have the same $N_{lag}=24$ lags $\ell=0,1,\cdots,23$.
Figure \ref{fi:actual_forecast} illustrates the fact that these errors do not change sign very often and may exhibit a strong temporal correlation.

\vskip 4pt
The removal of a trend and a seasonal component described in point 1) above is done in the same way for each of the $N^W_{zone}\times N_{lag}=5\times 24=120$ time series of lengths $N=17520$ time stamps for the historical data covering the two year period. As for the step 2) above, the marginal distributions exhibit heavy tails even if, contrary to what we observed in the case of loads, the lower tail is most of the time heavier than the upper tail.

\begin{remark}
The presence of tails, let alone of heavy tails, may seem counter intuitive at first. Indeed, the power produced by a single wind station is a non-negative number bounded from above by the capacity of the wind far. So for a specific wind farm, the deviation should have a distribution with support contained in an interval $[-c,+c]$ where $c$ is the capacity of the generation asset. Our explanation for the presence of tails in the zonal wind power is based on the aggregation effect and the fact that the capacity varies from one asset to another.
\end{remark}

\begin{figure}[h]
\centerline{
\includegraphics[width=5cm,height=5cm]{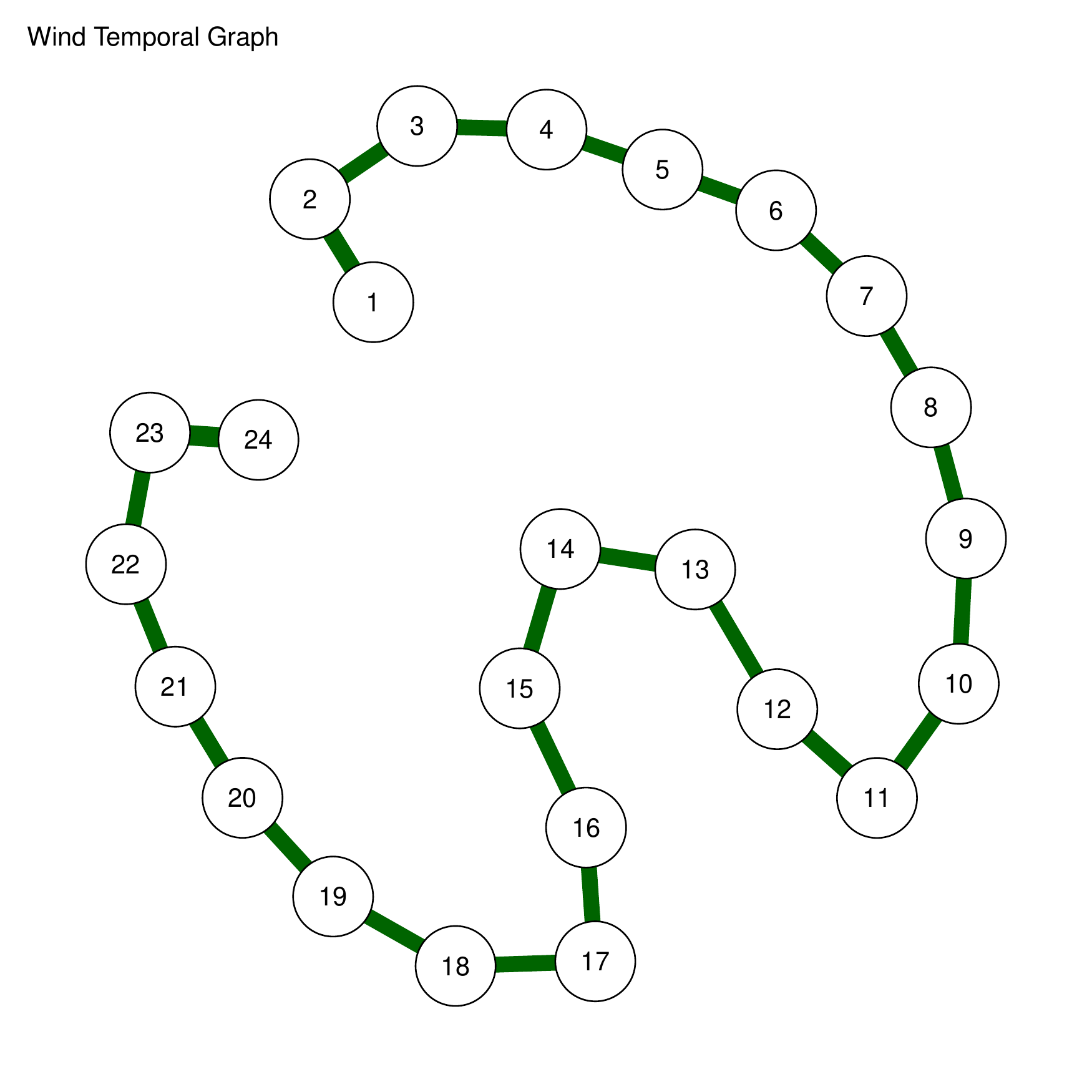}
\hskip 6pt
\includegraphics[width=5cm,height=5cm]{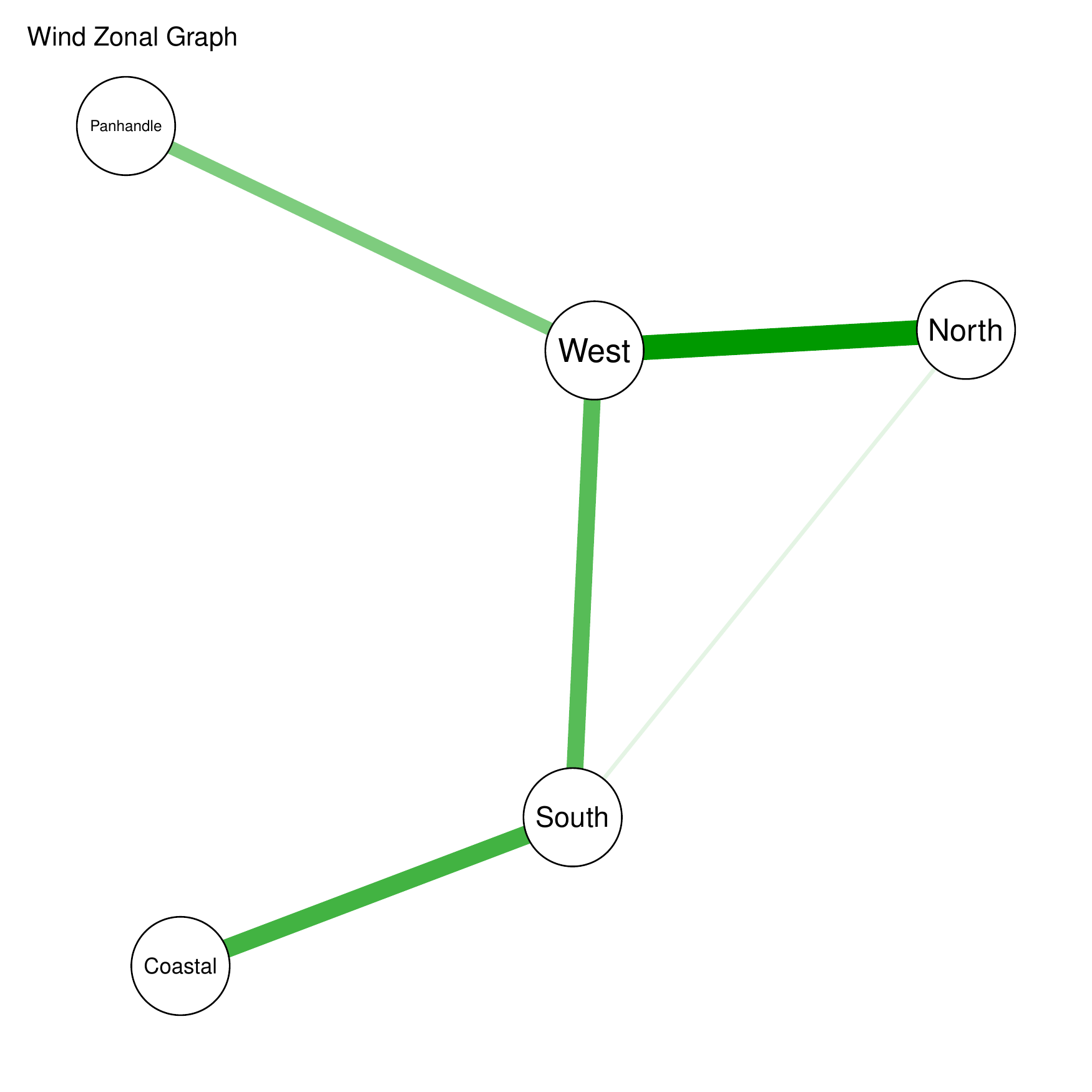}
\hskip 6pt
\includegraphics[width=5cm,height=5cm]{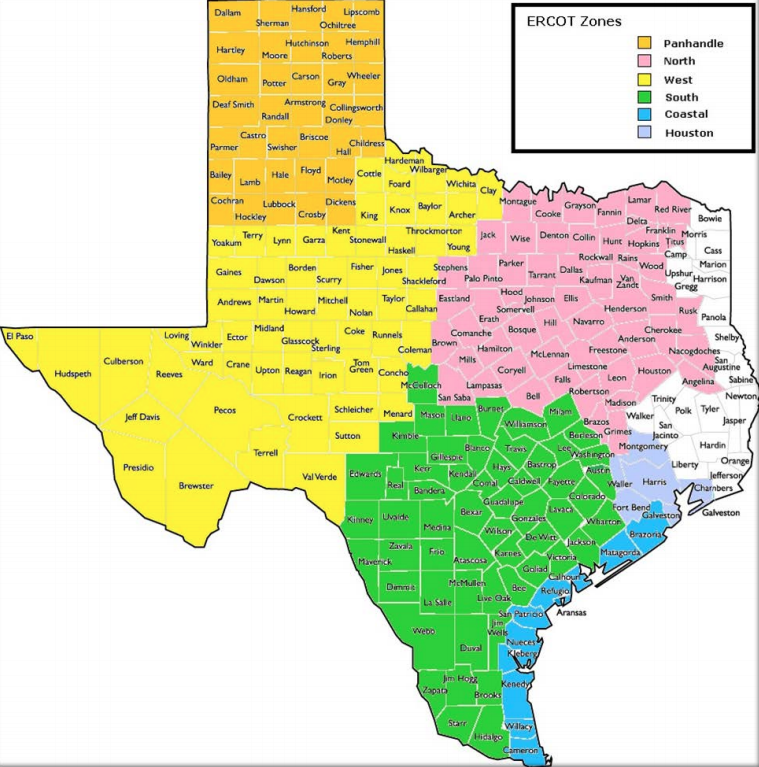}
}
\caption{Temporal component of the conditional correlation (left), spatial component of the conditional correlation (middle), geographical map of the five main wind power zones in ERCOT.}
\label{fi:wind_glasso}
\end{figure}

In any case, for each of the $120$ $(z,\ell)$ zone-lag couples, we fit a GPD $G_{z,\ell}^W$ to the  marginal distribution of the \emph{stationary} time series obtained by removing trend and seasonal component, we apply the cumulative distribution function of $G_{z,\ell}^W$ to each of the entries to produce a series with uniform marginals, and eventually standard Gaussian after applying the quantile function of the Gaussian distribution. This takes care of the steps 3) - 5) of the strategy used for the load data. Fitting a Gaussian graphical model like we did in step 6) above with the {\tt GEMINI} algorithm to capture the precision and covariance matrices leads to the analog of Figure \ref{fi:load_glasso} which we reproduce as Figure \ref{fi:wind_glasso}. Similar conclusions hold, the only significant difference being the stronger hint toward a Markovian temporal dependence structure given by the clear chain pattern of the temporal dependency graph. As before, the spatial dependency graph is consistent with the geographical locations of the wind power zones. Remember that we now have $5$ zones.

\begin{figure}[h]
\centerline{
\includegraphics[width=12cm,height=7cm]{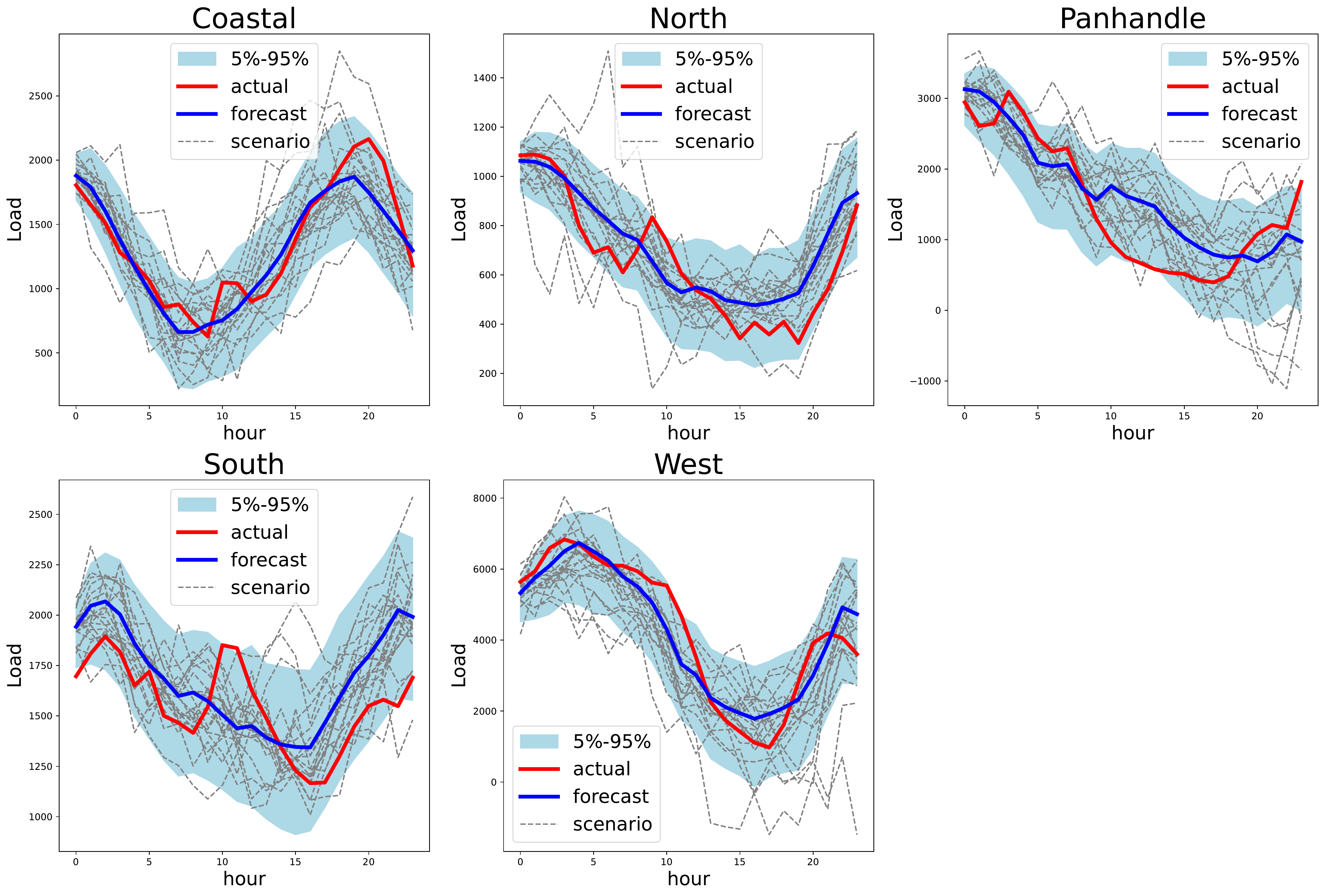}
}
\caption{Examples of a set of $1000$ Monte Carlo scenarios for the next $24$ hours of the actual wind power generation for July 1, 2018  at midnight. The red line gives the actual power observed over these $24$ hours, the blue line shows the point forecasts for these $24$ hours while the dashed lines are $20$ randomly selected scenarios. The gray bands are the trace left by the Monte Carlo scenarios after we remove the smaller $5\%$ and the larger $5\%$ of the bunch.}
\label{fi:wind_scenarios}
\end{figure}

Exactly as in the case of the load simulations, we can generate Monte Carlo scenarios for the actual wind power over the period spanned by $24$ hourly lags starting from a day-time $(d,l)$ by first generating Gaussian vector scenarios from the graphical Gaussian model, and then, reinstating the heavy tail GPD and adding the seasonal and trend components. Figure \ref{fi:wind_scenarios} gives an example. 

\section{\textbf{Modeling \& Simulating Jointly Electricity Loads and Wind Power}}
\label{se:altogether}

\begin{figure}[h]
\centerline{
\includegraphics[width=6cm,height=6cm]{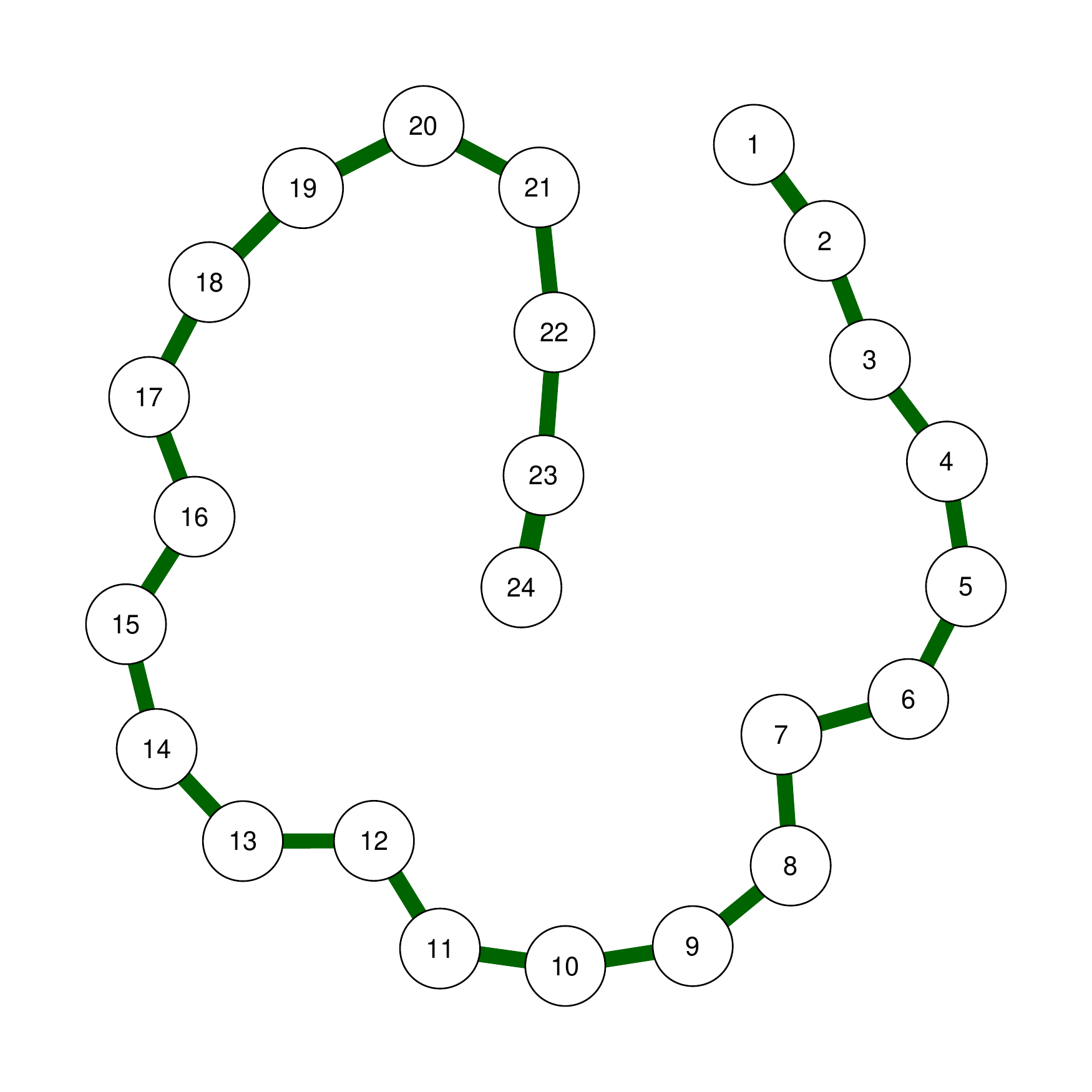}
\hskip 6pt
\includegraphics[width=6cm,height=6cm]{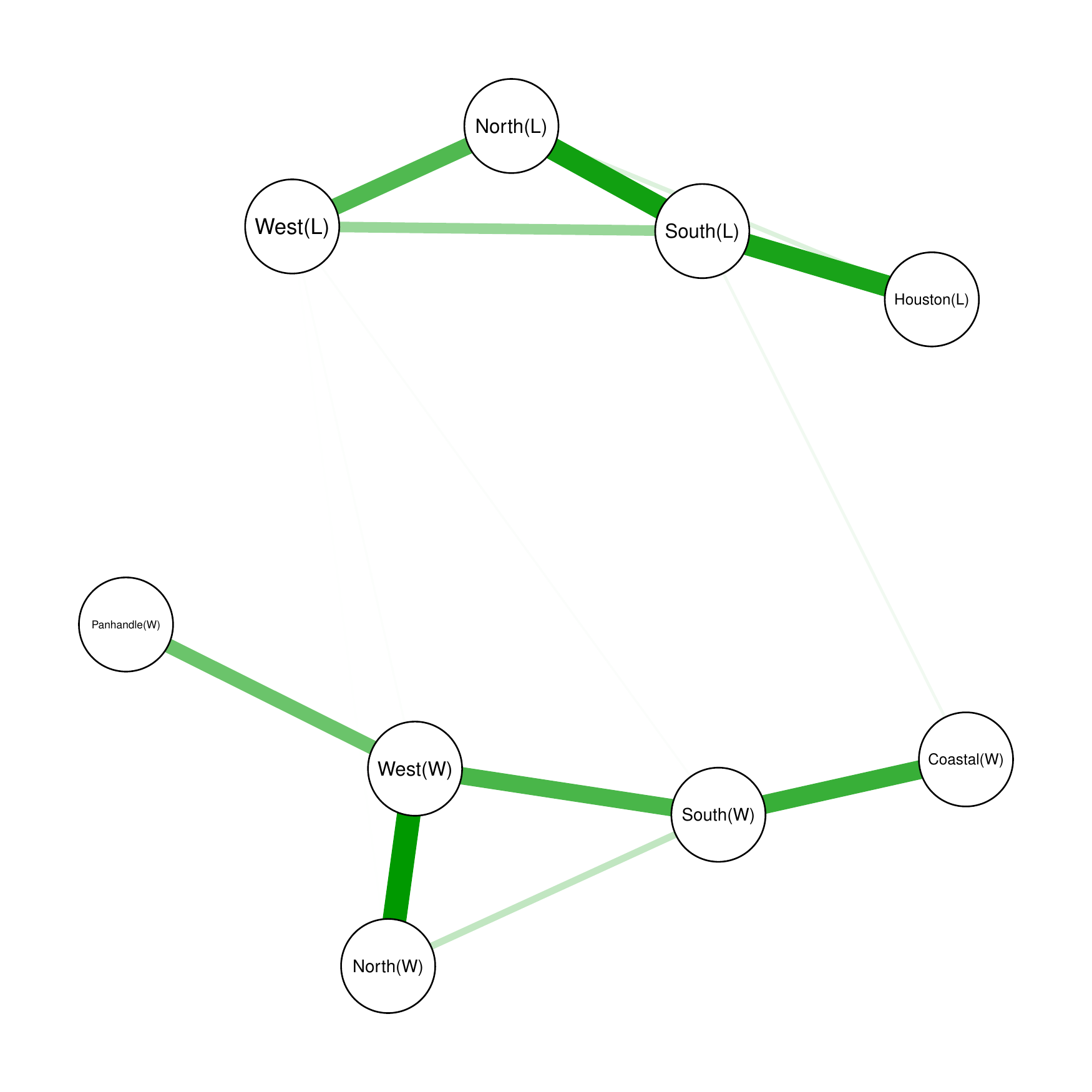}
}
\caption{Temporal component of the conditional correlation (left) and spatial component of the conditional correlation (right) for the joint analysis of the load deviations over the $4$ load zones and the $5$ wind power zones in ERCOT.}
\label{fi:altogether}
\end{figure}

This section will be even shorter than the previous one as the various steps of our modeling and simulation strategy should be clear by now. We consider for each day-hour ($d,h)$ in our historical data, the $N_{lag}(N^L_{zone}+N^W_{zone})=24(4+5)=216$-dimensional vector containing the $24$ hour lag-ahead, and for each of such a lag $\ell\in\{0,1,\cdots,23\}$ the numerical values of the load deviations in each of the $4$ load zones, and the $5$ numerical values of the wind power deviation remainders for each of the $5$ wind zones. All of the steps of the analysis strategy have already being taken, except for the graphical Gaussian model which is now $216$ dimensional. Applying the {\tt GEMINI} version of the {\tt glasso} estimation of the precision and covariance matrices we find the results reported in Figure \ref{fi:altogether}.

The graph underlying the temporal dependencies is again a linear chain as expected. This is consistent with the exponential temporal correlation used in many papers in the existing literature. However, the pattern appearing in the spatial dependency graph is striking. 
This graph is divided in two completely disconnected components, one component corresponding to the load zones, and the other to the wind power zones, each component giving an eerie incarnation of the graphs we obtained when we studied these factors separately.
This is a clear indication that load and wind power are uncorrelated, and independent at the Gaussian level.

\section{\textbf{Conclusion}}
\label{se:conclusion}
In this paper, we propose an original model for electricity demand and wind power which can be easily implemented to produce joint Monte Carlo simulations for the purpose of unit commitment and economic dispatch performed by highly sophisticated optimization programs. The novelty of our approach resides on 1) checking for the presence of heavy tails in the marginal distributions and the use of  generalized Pareto distributions to \emph{Gaussianize} the data in the spirit of the use of Gaussian copulas; 2) using graphical Gaussian model and producing precision matrix proxies with the {\tt gemini} variation on the {\tt glasso} algorithm, disentangling the temporal and spatial contributions to the correlations, and allowing for implementations with high dimensional state variables. We illustrate the  potential of the model and its implementation on data publicly available from the Texas Independent System Operator (ISO) website.

\section{\textbf{Declaration of Competing Interest}}
The authors declare that they have no known competing financial interests or personal relationships that could have appeared to influence the work reported in this paper.

\section{\textbf{Acknowledgments}}
Both authors were partially supported by ARPA-E grants DE-AR0001289 and DE-AR0001390 under the PERFORM program of the US Department of Energy. A preliminary version of the paper was presented in November 2020 during the INFORMS Annual Meeting, and on December 21, 2020 at a DOE Review Meeting. We would like to thank Mike Ludkovski (University of California at Santa Barbara), Ronnie Sircar (Princeton University) and GLen Swindle (Scoville Risk Partners) for enlightening conversations on the content of the paper.

\bibliographystyle{plain}

\end{document}